\documentclass[pageno]{jpaper}

\usepackage[normalem]{ulem}

\usepackage{graphicx} 
\usepackage{graphicx} 
\usepackage{subfig}
\usepackage{booktabs}
\usepackage{tabularx} 
\usepackage{multirow}
\usepackage{adjustbox} 
\usepackage{framed} 
\usepackage{float} 
\usepackage{stfloats} 
\usepackage{caption} 
\usepackage{mathptmx} 
\usepackage{amsmath}
\usepackage{fancyhdr}
\usepackage[bottom]{footmisc} 
\usepackage{lipsum}
\usepackage{tikz} 
\usepackage{xcolor} 
\usepackage{enumitem} 
\setlist{nolistsep}
\usepackage[normalem]{ulem}

\usepackage{hyperref} 
\hypersetup{colorlinks,allcolors=black}
\usepackage{amssymb}
\usepackage{pifont}
\usepackage{ifthen} 
\usepackage{xspace}
\usepackage{listings}
\usepackage{comment}

\usepackage[ruled,vlined,linesnumbered]{algorithm2e} 
\usepackage[noend]{algpseudocode} 
\algrenewcommand\algorithmicindent{1.5em} 

\usepackage{etoolbox}
\usepackage{soul} 
\sethlcolor{green}

\newcommand{\abb}[1]{{\textsf{\small{#1}}}\xspace}

\definecolor{cadmiumgreen}{rgb}{0.0, 0.42, 0.24}
\definecolor{carnelian}{rgb}{0.7, 0.11, 0.11}

\newcommand{\name}{gpu-let\xspace}
\newcommand{\Name}{Gpu-let\xspace}
\newcommand{\names}{gpu-lets\xspace}
\newcommand{\alg}{elastic partitioning\xspace}
\newcommand{\Alg}{Elastic partitioning\xspace}
\renewcommand{\paragraph}[1]{\vspace{0.2ex}\noindent{\bf #1}}

\newcommand{\xmark}{\textcolor{carnelian}{\ding{55}}}
\newcommand{\greencheck}{{\textcolor{cadmiumgreen}{\checkmark}}}
\newcommand{\redx}{{\textcolor{red}{\xmark}}}

\newboolean{color}
\setboolean{color}{true}

\definecolor{mGreen}{rgb}{0,0.6,0}
\definecolor{mGray}{rgb}{0.5,0.5,0.5}
\definecolor{mPurple}{rgb}{0.58,0,0.82}
\definecolor{backgroundColour}{rgb}{0.95,0.95,0.92}
\lstdefinestyle{CStyle}{
    backgroundcolor=\color{backgroundColour},
    commentstyle=\color{mGreen},
    keywordstyle=\color{magenta},
    numberstyle=\tiny\color{mGray},
    stringstyle=\color{mPurple},
    basicstyle=\footnotesize,
    breakatwhitespace=false,
    breaklines=true,
    captionpos=b,
    keepspaces=true,
    numbers=left,
    numbersep=5pt,
    showspaces=false,
    showstringspaces=false,
    showtabs=false,
    tabsize=2,
    language=C
}
\newcolumntype{M}[1]{>{\centering\arraybackslash}m{#1}}

\begin{document}
\title{Multi-model Machine Learning Inference Serving with GPU Spatial Partitioning}
\date{}

\author{Seungbeom Choi \quad Sunho Lee \quad Yeonjae Kim \quad Jongse Park \quad Youngjin Kwon \quad Jaehyuk Huh \\School of Computing, KAIST}
\maketitle

\thispagestyle{empty}

\begin{abstract}

As machine learning techniques are applied to a widening range of applications, high throughput machine learning (ML) inference servers have become critical for online service applications. Such ML inference servers pose two challenges: first, they must provide a bounded latency for each request to support consistent service-level objective (SLO), and second, they can serve multiple heterogeneous ML models in a system as certain tasks involve invocation of multiple models and consolidating multiple models can improve system utilization. To address the two requirements of ML inference servers, this paper proposes a new ML inference scheduling framework for multi-model ML inference servers. The paper first shows that with SLO constraints, current GPUs are not fully utilized for ML inference tasks. To maximize the resource efficiency of inference servers, a key mechanism proposed in this paper is to exploit hardware support for spatial partitioning of GPU resources. With the partitioning mechanism, a new abstraction layer of GPU resources is created with configurable GPU resources. The scheduler assigns requests to virtual GPUs, called gpu-lets, with the most effective amount of resources. The paper also investigates a remedy for potential interference effects when two ML tasks are running concurrently in a GPU. Our prototype implementation proves that spatial partitioning enhances throughput by 102.6\% on average while satisfying SLOs.


\end{abstract}

\section{Introduction}

The wide adoption of machine learning (ML) techniques poses a new challenge in server system designs. Traditional
server systems have been optimized for CPU-based computation for many decades. However, the regular
and ample parallelism in widely-used deep learning algorithms can exploit abundant parallel execution units
in GPUs. Although powerful GPUs have been facilitating the training computation of deep learning models,
the inference computation is also moving to the GPU-based servers due to the increasing computational
requirements of evolving deep learning models with deeper layers~\cite{nvidia:inference, sosp:nexus, atc:mark}.

However, the GPU-based inference servers must address different challenges from the batch-oriented
processing in ML training servers. First, inference queries must be served within a bounded time to
satisfy service-level objectives (SLOs). Therefore, not only the overall throughput of systems is
important, but bounded response latencies for processing inference queries are also critical to maintain 
consistent service quality~\cite{nsdi:clipper, sosp:nexus, atc:mark}.

Second, to improve the utilization of server resources, multiple heterogeneous models need to
be served by a single node. As the amount of memory in GPUs increases, each GPU can maintain
the parameters for multiple models in GPU DRAM, which enables fast switching of ML models without
swapping models out of GPU. As even a single service can include multiple heterogeneous ML models~\cite{sosp:nexus},
multiple models with different purposes co-exist in a system.
The heterogeneity of ML models raises new scheduling challenges to map concurrent requests of heterogeneous
models to multiple GPUs. Incoming queries for different ML models with their own
computational requirements, must be properly routed to
the GPUs to meet the SLO, while improving the overall throughput.

While the demands for GPU-based ML inferences have been growing, the computational capability of GPUs based
on parallel execution units has been improving precipitously. Such ample parallel execution units
combined with increasing GPU memory capacity allow multiple ML models to be served by a single GPU.
In the prior study~\cite{sosp:nexus}, more than one model can be mapped to a GPU, as long as the GPU can
provide the execution throughput to satisfy the required SLO.
However, unlike CPUs which allow fine-grained time sharing with efficient preemption,
GPUs perform only coarse-grained kernel-granularity context switches.
In addition, to reduce the variance of execution latencies for a request, the GPU is allocated for
a single batch of requests for a given model before accepting the next batch of the same or a different model~\cite{sosp:nexus}.
Such coarse-grained time sharing incurs inefficient utilization of enormous computational capability of GPUs, as
a single batch of an ML inference may not fill the entire GPU execution units.

However, the recent advancement of GPU architecture opens a new opportunity to use abundant execution resources of GPUs.
Recent GPUs support an efficient spatial partitioning of
GPUs resources (called MPS mechanism~\cite{nvidia:MPS}). The MPS mechanism after the NVIDIA Volta architecture
supports that the computational resources of a GPU can be partitioned to run different contexts simultaneously.
Such a unique spatial partitioning mechanism can augment the limited coarse-grained
time sharing mechanism, as the GPU resource can be spatially partitioned to serve
multiple ML tasks concurrently.
This unique spatial and coarse-grained temporal resource allocation in GPUs
calls for a novel abstraction to represent partitioned GPUs and a new scheduling framework
targeting high throughput ML servers under SLO constraints.

To address the emerging challenges of ML scheduling in partitionable GPUs, this
paper proposes a new abstraction for GPUs called \textit{\name}, which can create multiple virtual GPUs out of a single physical
GPU with spatial partitioning. 
The new abstraction can avoid the inefficiency of the coarse-grained time sharing, by creating and assigning
the most efficient GPU share for a given ML model. Each \name can be used for ML inferences independently from other tasks.
Such a new abstraction of GPU resources allows the predictable latencies for ML execution even when multiple
models are concurrently running in a GPU, while achieving improved GPU utilization.

Based on the \name concept,
we propose a ML inference framework prototyped on PyTorch interface. It can serve concurrent heterogeneous ML models in multi-GPU environments.
The scheduling framework implements the \name  abstraction based on both spatial and temporal
shares of GPU resources. 
For each ML model, its computational characteristics are measured and registered to the framework.
Based on the pre-profiled information of the ML models, the scheduler routes requests to where the throughput would be
maximized, while satisfying the SLO constraints.

To allocate the spatial and temporal GPU share, the framework creates \names representing
all or part of GPU resources using the MPS support. 
One necessary mechanism for the spatial and temporal partitioning of GPU shares is to identify the potential
performance overheads when two models are concurrently running on a GPU, where two virtual GPUs are mapped.
Although a prior study modeled such interference of concurrent kernels in pre-Volta MPS systems, this study proposes
a newly tuned interference estimation model for concurrent kernel executions of ML inference workloads.

We evaluated the proposed ML inference framework on a four-GPU server system. 
Each physical GPU can provide up-to two virtual \names with adjustable capability.
The evaluation with the four-GPU server shows that the proposed scheduling technique with \names can improve
the throughput with SLO constraints for five ML inference scenarios by 102.6\%, compared to the one without partitioning GPU resources.
The source codes will become publicly available after publication. 

This study explores a new resource provisioning space of GPUs for machine learning inference serving. The contributions of this paper are as follows:

\begin{itemize}
\item This paper identifies that with SLO constraints, ML models commonly cannot fully utilize current high performance GPUs. 
      Due to the SLO constraints, only limited batch-processing is possible for a model, which cannot fill the entirety of execution units 
      in a GPU. 
\item It proposes a new GPU abstraction for GPU-based ML inference servers, named \textit{gpu-let}, to support virtual GPUs with partitions of resources 
      out of physical GPUs.
\item It proposes a scheduling framework for \names, which enables the effective utilization of GPUs. The framework schedules tasks with a scalable algorithm.
\item It proposes an interference model for \names for concurrent ML inference execution on
      partitions of a single GPU.
\end{itemize}

The rest of the paper is organized as follows.
Section 2 describes the background of ML computation on GPUs and the prior scheduling technique for MLs on GPUs.
Section 3 presents the motivational analysis of heterogeneous ML tasks on multiple GPUs.
Section 4 proposes our design for \names to efficiently utilize GPU resources for heterogeneous ML tasks.
Section 5 describes the implementation details, and Section 6 presents the experimental results.
Section 7 presents the related work, and Section 8 concludes the paper.

\section{Background}
\label{sec:background}
\subsection{ML Inference Serving System}
\label{sec:inference-system}
%
%
%
%
%
While many ML inference serving system still employ CPUs, the increasing number of service vendors are adopting GPUs \cite{vldb:rafiki, hpca:facebook18, sosp:nexus, osdi:gandiva, atc:mark, atc:poseidon, aws:sagemaker, nips:tensorflow_serving, nsdi:clipper, isca:djinn, nvidia:tensorRT} or even hardware accelerators such as TPUs \cite{isca:tpu, micro:planaria, hpca:prema, isca:jwkim}. 
This trend is attributed to the two primary reasons: 1) the heavy computational intensity of the ML models, and 2) the high parallelism and throughput that hardware accelerators offer.
While GPUs offer lower latency for ML inference compared to CPU systems, obtaining high utilization is a challenging task, unlike ML training.
%
The key difference between training and inference in terms of the GPU utilization is the suitability for \emph{batching}. 
%
%
%
For training, since the input data is ready, the system can batch any number of input data, which allows GPUs to effectively leverage the massive parallelism. 
%
In contrary, ML inference server is an on-demand system where once the inference requests arrive, then and only then, the inference tasks can be scheduled to the compute engines.
One scheduling option is to wait until the desirable number of inference requests to be accumulated and then initiate the execution for the large batch. 
However, the problem is, the applications cannot wait for the batch collection indefinitely, due to the service-level objective (SLO) requirements. 
Moreover, inference servers are multi-tenant systems where different models have different SLOs, which makes the SLO guarantees even more challenging.   
Thus, choosing batch sizes that maximize throughput without violating SLOs becomes the core challenge in the SLO-aware inference scheduling.

\subsection{SLO-oriented ML Inference Scheduling}
\label{sec:sbp}

ML inference scheduling problem on GPU-based multi-tenant serving systems resembles the traditional bin packing problem where the bins are GPUs and the items are inference requests.
Similar to the bin packing algorithm, the objective of inference scheduling is to serve the inference requests on a minimal number of GPUs, while satisfying the SLOs.
The bin's capacity constraints are the available resource on the GPUs, and the item weights are the necessary GPU resource to handle the given inference requests. 
The fundamental difference of our scheduling problem with the traditional bin packing is that ones can lower the item weight by choosing a larger batch size as long as the increased request latency still meets the SLO. 
This disparity formulates a unique scheduling problem.

An inspiring prior work, Nexus~\cite{sosp:nexus}, has tackled this problem and proposed a novel variant of bin packing algorithm, namely \textit{squishy bin packing} (SBP).
The term, ``squishy'', is originated from the property that the required resource for processing a request (i.e., item) and its latency vary as the batch size changes. 
The SBP algorithm takes as input a set of models, each of which comes with its request rate, assuming that the requests regularly arrive. 
The SBP algorithm finds the heuristically optimized mappings from models to GPUs such that the inference requests for the assigned models would not create the SLO violations under the given request rates. 
\begin{figure}[t]
	\centering
	\includegraphics[width=0.7\linewidth]{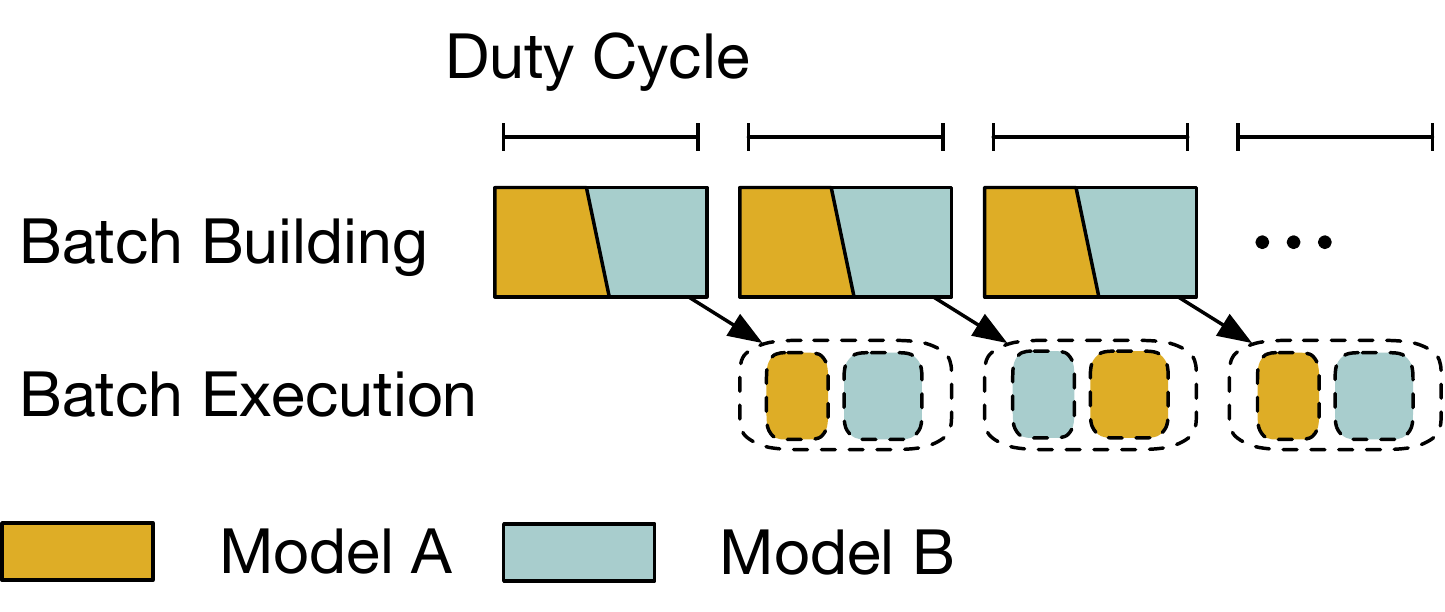}
	\caption{Round-based execution for two models consolidated on a GPU. The interval for a round is called ``duty cycle''.}
	\label{fig:scenario_overview}
\end{figure}
Figure~\ref{fig:scenario_overview} illustrates the example scenario on a serving server. 
In this scenario, the server is handling two consolidated models, A and B, building and executing the per-model batches simultaneously. 
The SLO violation occurs when the summation of batch building time and batch execution time exceeds either of the SLOs.
Therefore, the SBP algorithm heuristically finds a maximum possible duration for batch building and the corresponding batch sizes in such a way that all the consolidated models would not violate the SLOs.
The duration is called duty cycle.
The SBP algorithm repeats the duty cycles in a pipelined fashion until there is a change on the request rates, which would require a rescheduling. 
\begin{figure}[t]
	\centering
	\includegraphics[width=0.8\linewidth]{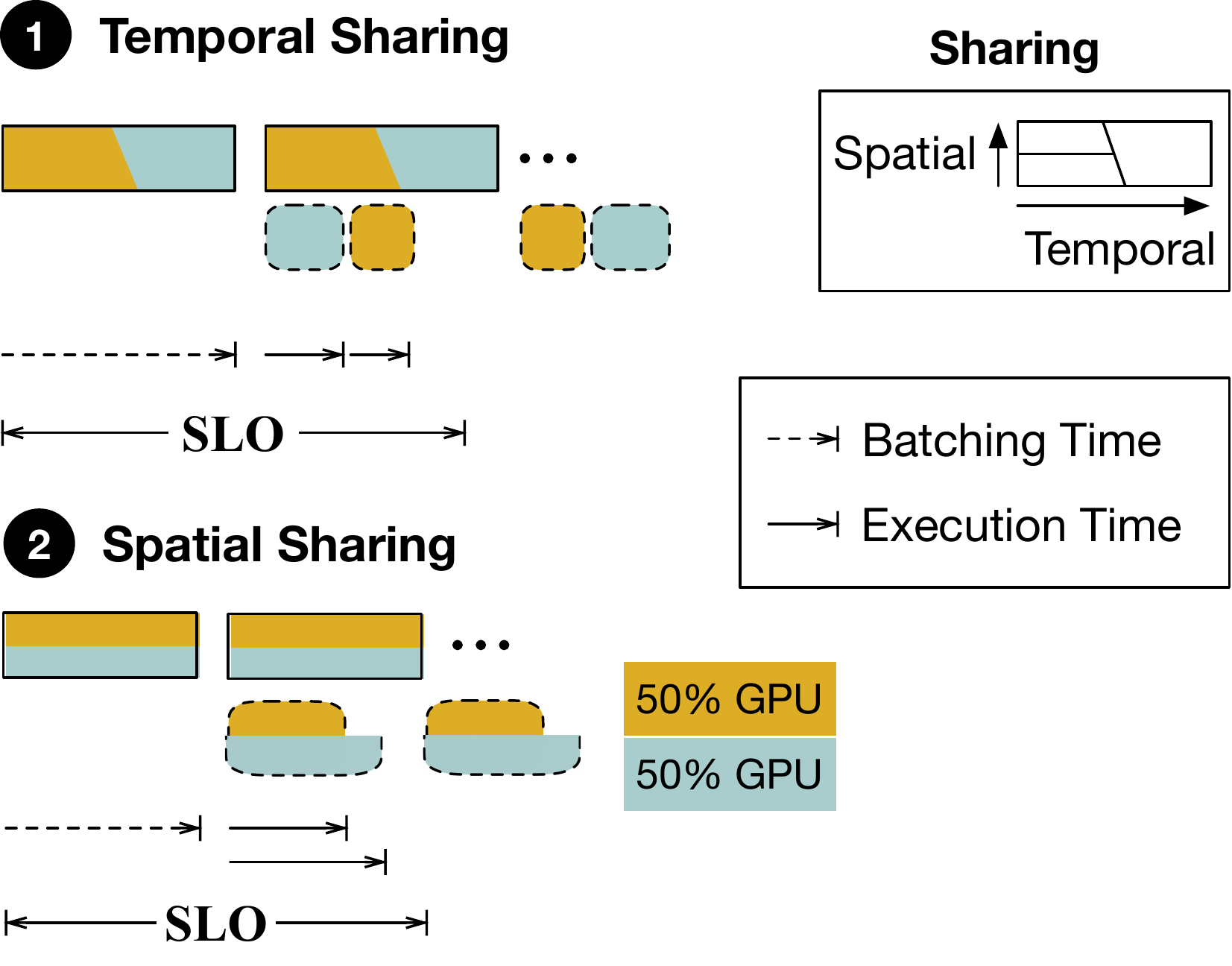}
	\caption{Temporal and spatial sharing.}
	\label{fig:time_spatial_sharing}
\end{figure}
To consolidate multiple inference executions on a server, there are two different approaches, as illustrated in Figure~\ref{fig:time_spatial_sharing}. 
Temporal sharing refers to the serialization of executions on GPUs where each inference takes up the entire GPU resource, while spatial sharing is a resource partitioning approach that splits GPU resource into multiple pieces and multiple inference executions are performed at the same time. 
The SBP algorithm employs the temporal sharing approach.
However, the temporal sharing may potentially cause resource underutilization when the batch size is insufficiently large to leverage the existing parallelism on GPUs. 
Instead, this paper aims to simultaneously employ the temporal and spatial sharing so as to better utilize the GPU resource and extract higher throughput, still without violating the SLOs. 

\subsection{Spatial Sharing on GPU}
\label{sec:mps}
Modern server-scale GPUs natively offer the spatial sharing feature in various forms depending on the generations of GPU microarchitectures. 
As NVIDIA is a leading and almost sole GPU vendor for the spatial sharing technology, we focus on the NVIDIA-provided technologies here.
NVIDIA offers Multi-Process Service (MPS) technology, which enables the CUDA runtime to 1) intercept the CUDA calls produced from possibly multiple GPU contexts, 2) consolidate the multi-context kernels as a single virtual context, and 3) invoke the collection of kernels as if there is only a single context. 
This way, MPS allows multiple contexts to be parallelized on GPUs for higher utilization, but since software has no control over how the kernel consolidation is done, the resource contention could lead to the high performance volatility. 
To address this limitation, the recent generations of NVIDIA, Volta and Ampere, have added the computation resource provisioning and memory bandwidth isolation features on top of their prior generations, respectively. 
With these resource partitioning features, the users can split the given resource of a GPU into a set of \textit{virtual} GPUs, each of which is assigned to a fraction of GPU resource, which we call \textit{gpu-let} in the rest of this paper\footnote{In this paper, we only use the computation resource provisioning technique since we have at our disposal 2080 Ti GPUs, the microarchitecture of which is Turing, an older generation than Ampere that offers the memory bandwidth isolation feature.}.  
Also, we will call the GPU resource splitting as \textit{GPU partitioning}. 
%



\section{Motivation}

%
%
\subsection{Implications of Batching and GPU Partitioning}
\label{sec:perf-impl}
To better understand the performance implications of batching and GPU partitioning, we perform an experimental study, using four ML models: GoogLeNet, ResNet50, SSD-MobileNet-V1, and VGG-16. 
The detailed descriptions for the ML models and GPU server specifications are provided in Section~\ref{sec:method}.

\begin{figure}[t]
  \centering
  \captionsetup[subfloat]{captionskip=0.3ex, labelformat=empty} 
  \hspace{1.5em}\subfloat[]{\includegraphics[scale=0.14]{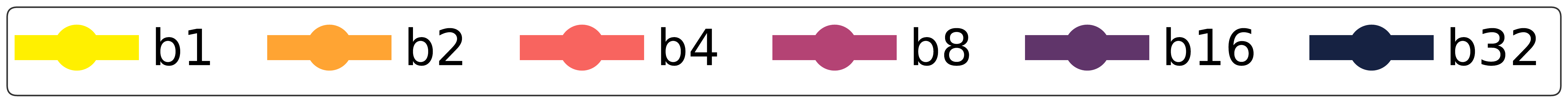}}\\[-4ex]
  \subfloat[(a) GoogLeNet]{\includegraphics[scale=0.066]{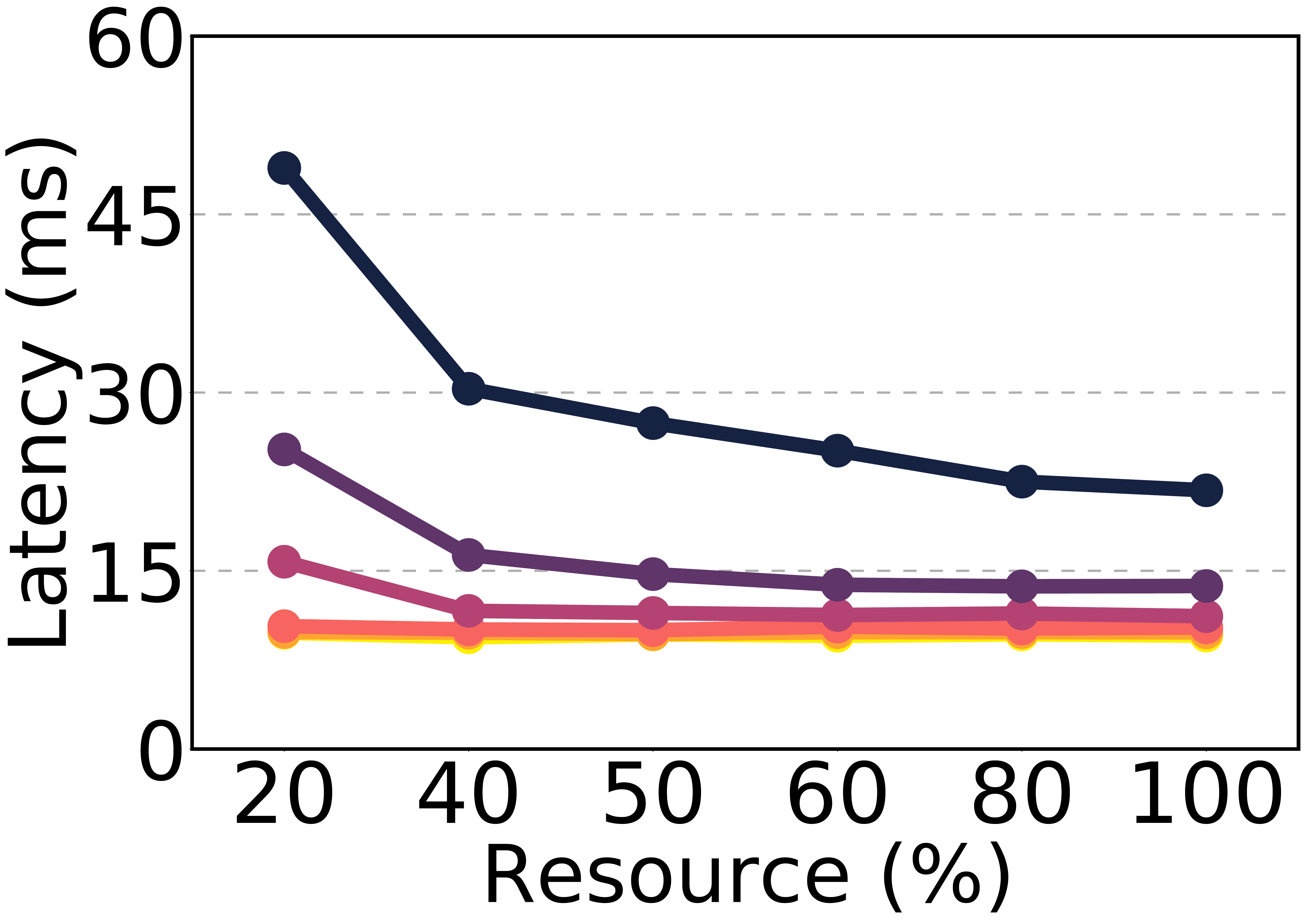}}\hfill
  \subfloat[(b) ResNet50]{\includegraphics[scale=0.066]{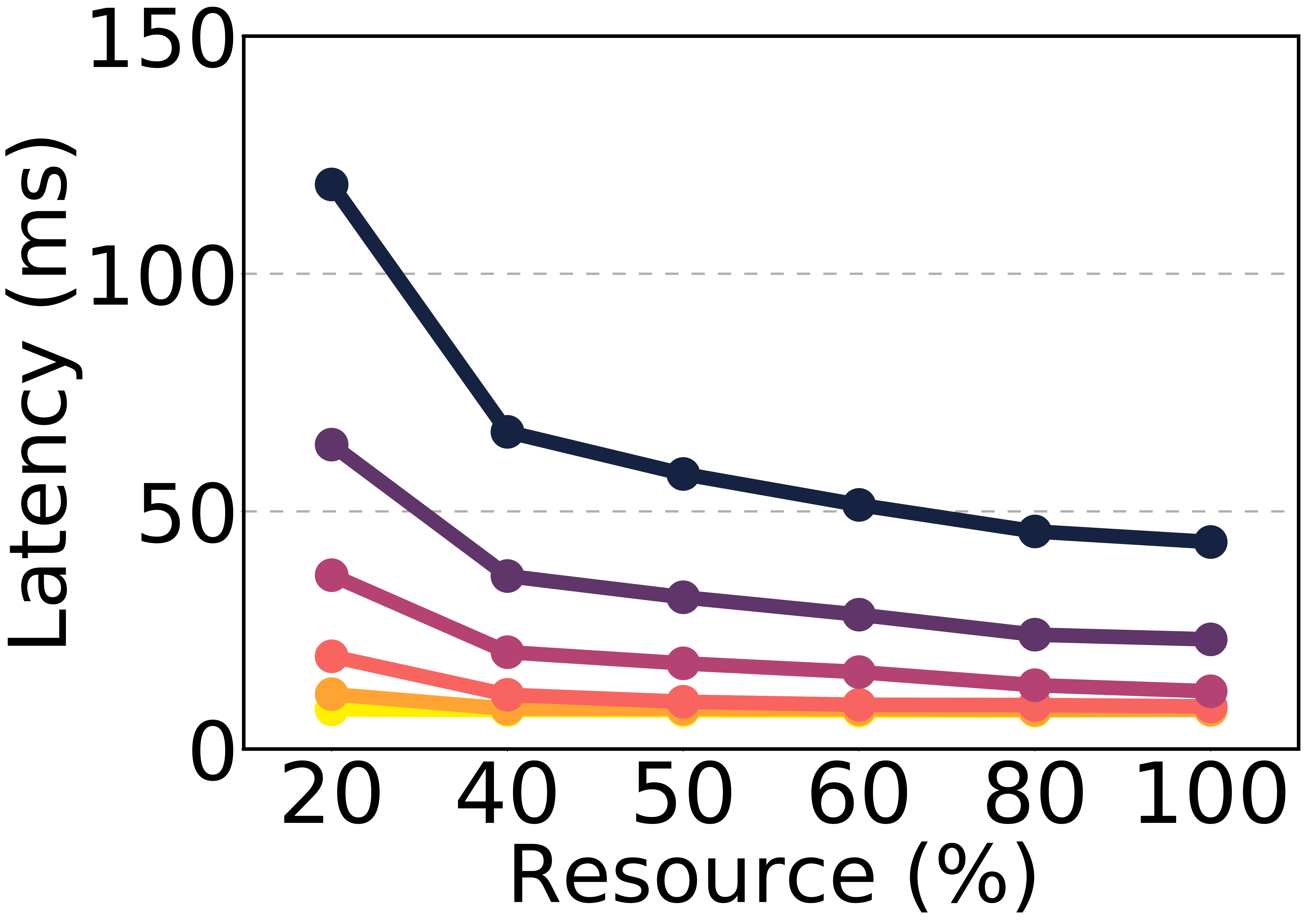}}\\[-2.0ex]
  \captionsetup[subfloat]{captionskip=+0.3ex}
  \subfloat[(c) SSD-MobileNet-V1]{\includegraphics[scale=0.066]{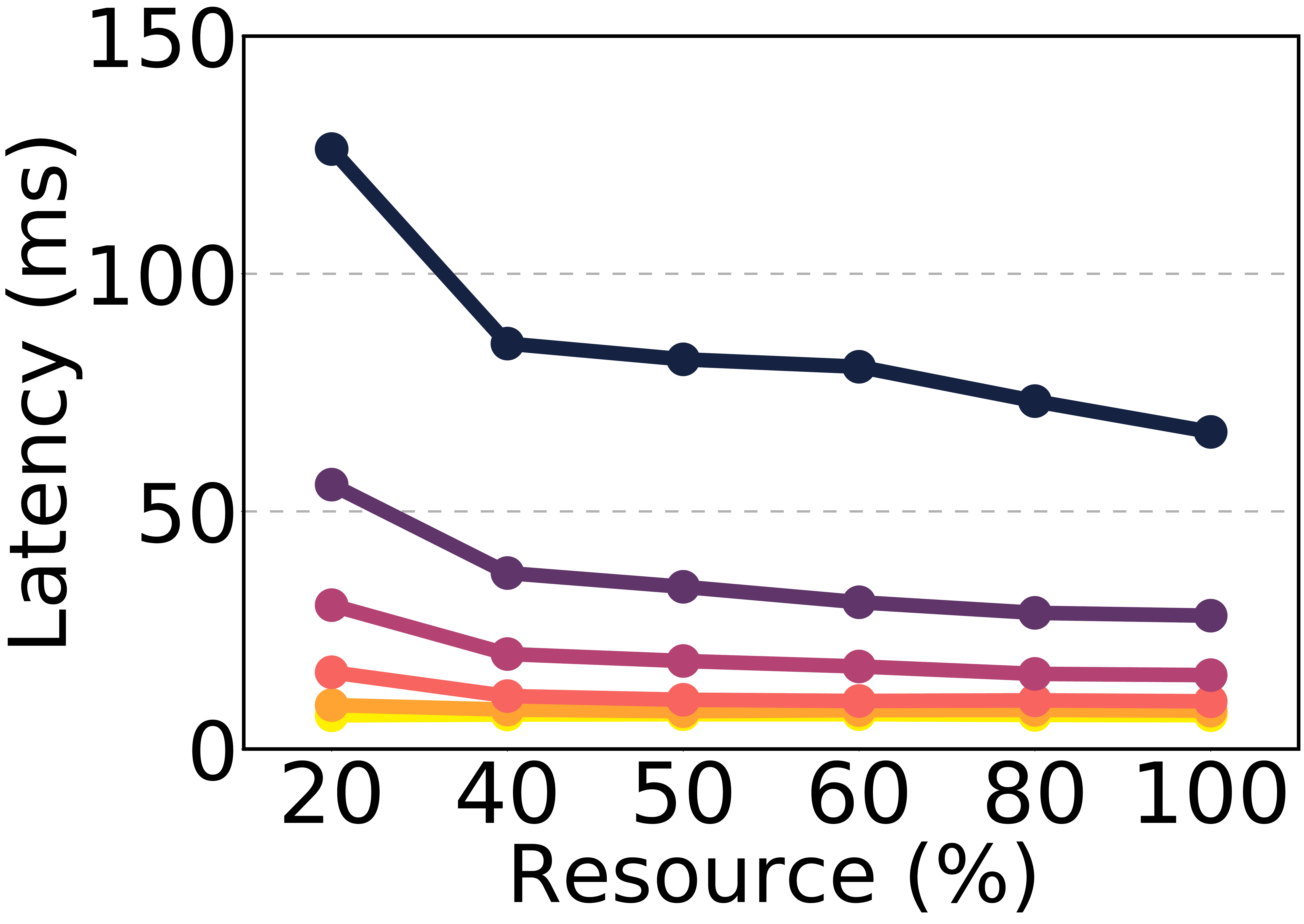}}\hfill
  \subfloat[(d) VGG-16]{\includegraphics[scale=0.066]{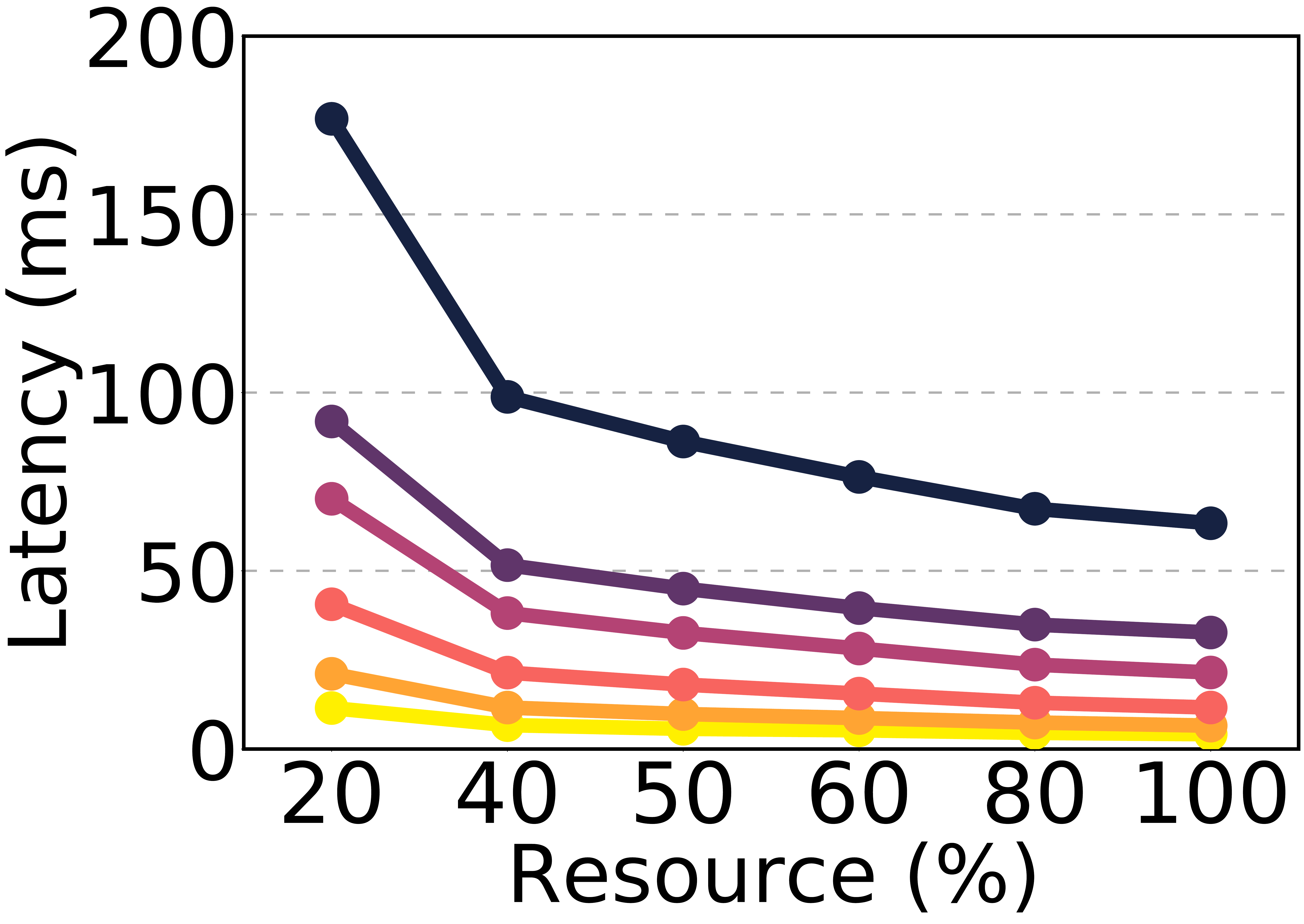}}
  \caption{Batch inference latency as the fraction of compute resource assigned to the model inference changes from 20\% to 100\%, for the four ML models.}
  \label{fig:motiv-batch-part}
\end{figure}
\paragraph{Implications on latency:}
Figure~\ref{fig:motiv-batch-part} shows the batch inference latency results as the batch size increases from 1 to 32. 
For each batch size, we sweep through the increasing fractions of GPU compute resource (i.e., gpu-let size), which range from 20\% to 100\% to observe how efficiently the inference execution utilizes the additional compute resource.  
%
%
The large slope of the curves implies that the inference execution for the particular batch size can use the additional resource effectively to reduce the latency. 
However, the flat region reflects that the additional resource is not helpful and thus wasted.   
Hence, selecting the gpu-let placed at the curve knee is considered to be the most cost-effective choice. 
When the batch size is large, the latency significantly drops as more resource is added. 
In contrary, with the small batch, the latency is not largely affected by the compute resource size, which implies the resource underutilization. 
These results suggest that exclusively occupying the entire GPU for an inference execution may be suboptimal and open novel opportunities for efficient ML serving systems through spatial GPU resource sharing.  

%


\paragraph{Implications on schedulability:}
When incoming request rates are beyond the level that the inference server can cope with, the inference server has no choice but to return ``Not Schedulable'' since the SLO will not be met.
Therefore, by improving the GPU utilization and in turn achieving higher throughput, a scheduler may be able to improve the \textit{schedulability}, which is a helpful metric to examine the effectiveness of scheduling algorithms for the SBP problem.

To evaluate the potentials of GPU partitioning on the schedulability improvement, we populate a significantly large number of possible multi-tenant inference serving scenarios and then measure the schedulability by counting the number of \textit{schedulable} ones among the scenarios. 
For each scenario, a model is associated with a request rate among the following four: 0, 200, 400, and 600 requests per second (req/s).
%
Excluding the scenario where all the models have zero request ratio, we use the total of 1,023 (= $4^5-1$) scenarios for the experiment.
%


\begin{figure}[t]
	\begin{center}
		\includegraphics[width=1.0\linewidth]{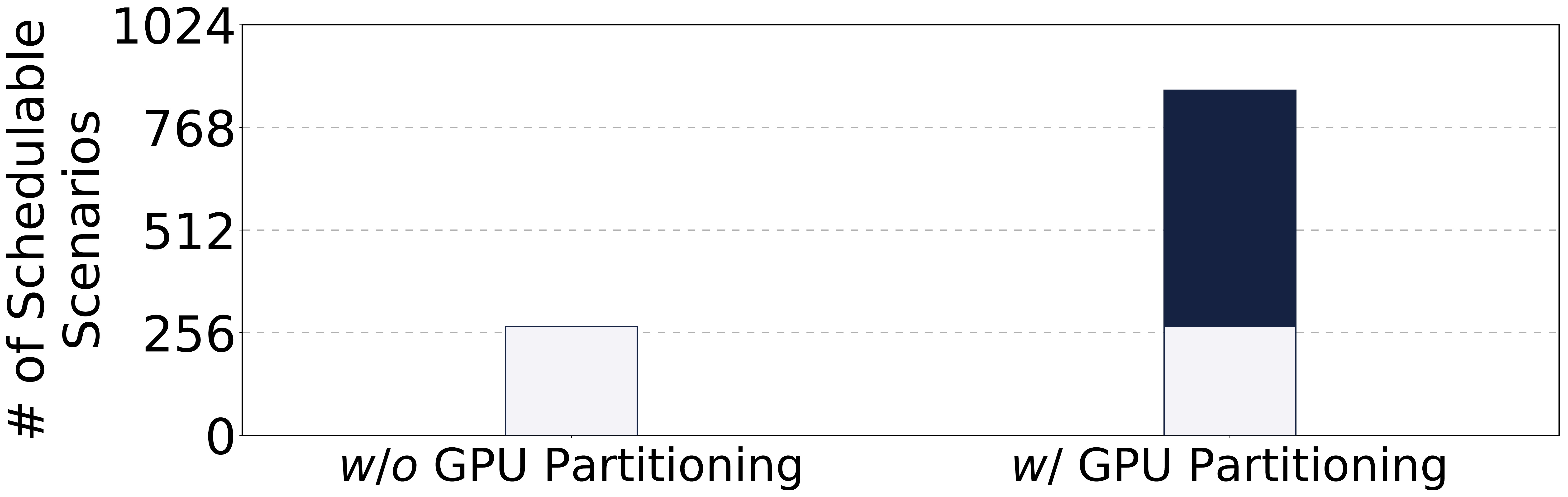}
	\end{center}
	\vspace{-2ex}
	\caption{Number of schedulable scenarios when the SBP algorithm performs the scheduling \textit{without} (right) and \textit{with} (left) the GPU partitioning support.}
	\label{fig:success-case}
\end{figure}

\begin{figure}[t]
  \centering
  \captionsetup[subfloat]{captionskip=0.3ex, labelformat=empty}
  \subfloat[]{\includegraphics[scale=0.145]{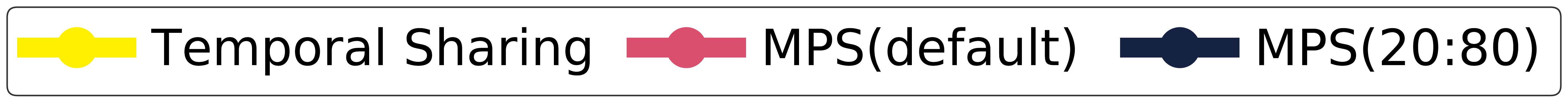}}\\[-5ex]
  \subfloat[(a) LeNet]{\includegraphics[scale=0.135]{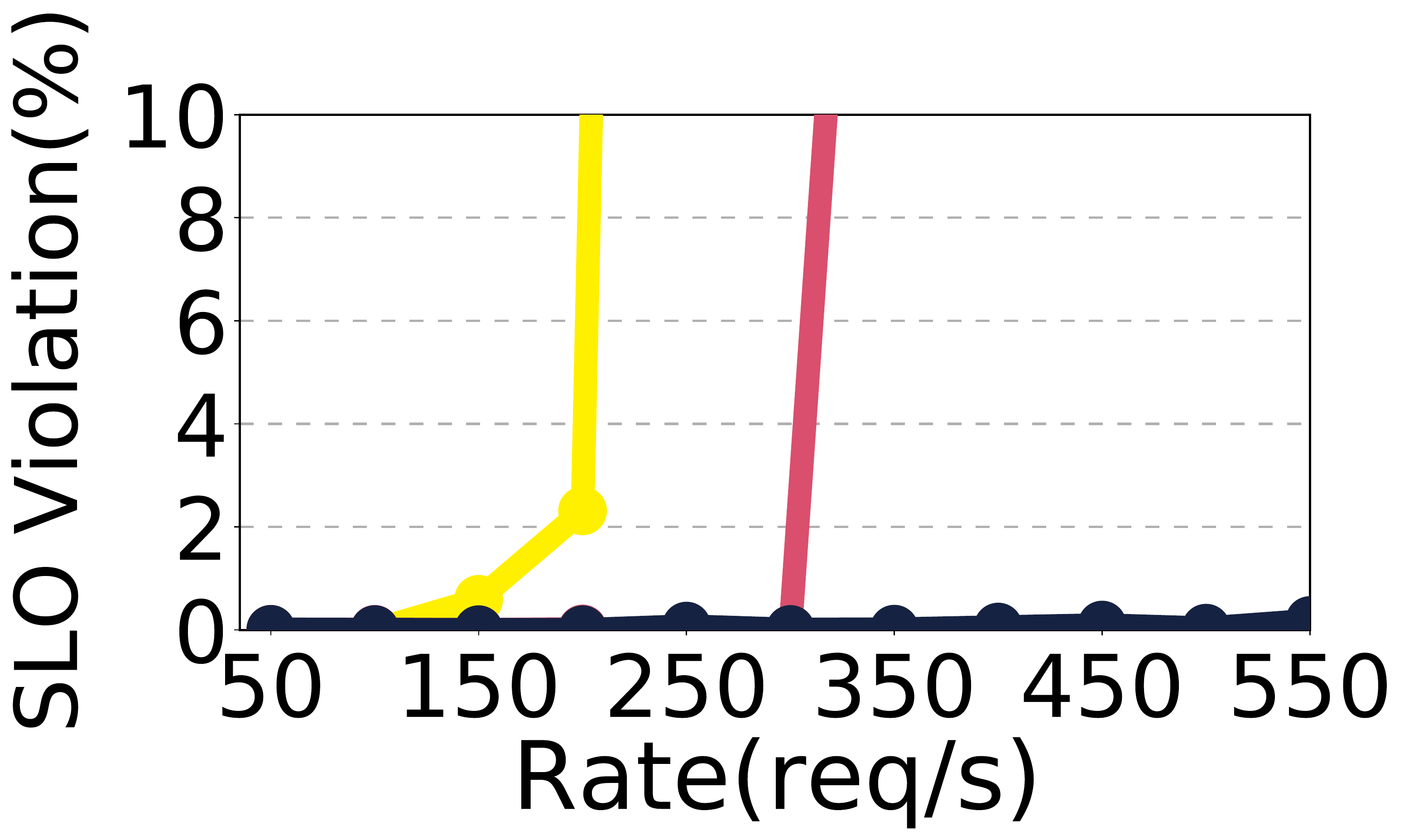}}\hfill
  \subfloat[(b) VGG-16]{\includegraphics[scale=0.135]{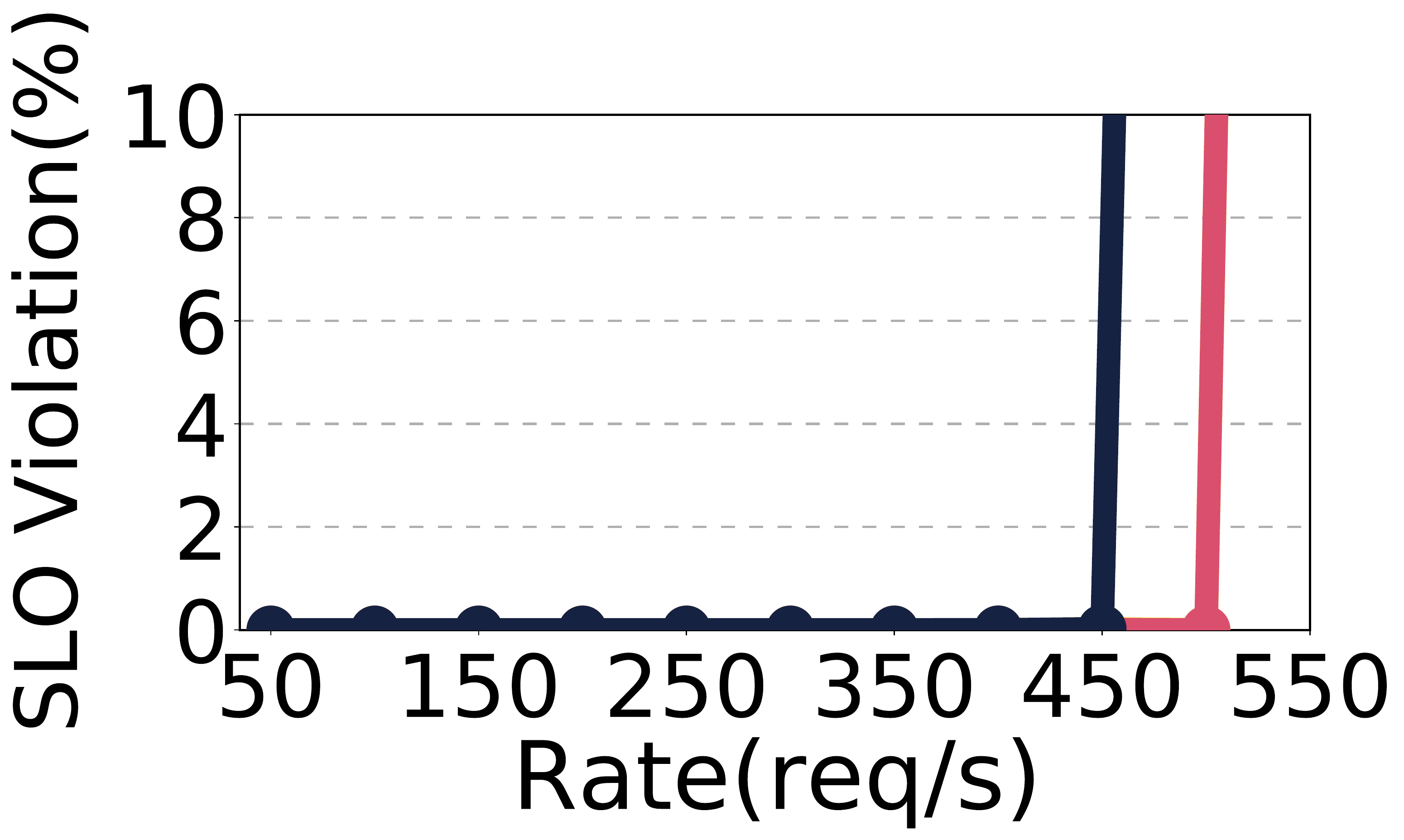}}\\[-1.0ex]
  \caption{SLO violation of LeNet and VGG-16 when they are consolidated using three different resource sharing schemes. MPS(20:80) refers to two gpu-lets with 20\% and 80\% resource.} 
  \label{fig:slo-violation}
\end{figure}

%
Figure~\ref{fig:success-case} reports the number of schedulable scenarios when we use the two different scheduling algorithms: 1) the default SBP algorithm without GPU partitioning support, and 2) SBP algorithm that independently schedules two evenly split gpu-lets. 
%
%
With GPU partitioning, most scenarios identified as \textit{un}schedulable by the baseline scheduling algorithm are eliminated. 
%
Note that for our constrained experiment setup, we could completely eliminate unschedulable scenarios, yet it is not readily generalizable. 
The results imply that GPU partitioning is potentially capable of putting wasted GPU compute power to use, enabling higher resource utilization. 
For deeper investigation, we look into a specific case scenario where LeNet and VGG-16 are consolidated on 20\% and 80\% gpu-lets (i.e., MPS(20:80)), respectively. 
Figure~\ref{fig:slo-violation} presents the SLO violation results as we raise the request rates for the two models with the same degree. 
When the two models are consolidated using the default temporal sharing (i.e., Temporal-Sharing) and MPS-based spatial sharing without static resource partitioning (i.e., MPS(default)), the SLO violation spikes up as we increase the incoming request rates. 
However, the MPS with static GPU partitioning can cope with significantly higher request rates. 
This result indicates that GPU partitioning could be useful to enable GPU to accommodate higher request rates by unlocking more throughput.

\subsection{Interference in Consolidated Executions}


GPU partitioning allows to enhance the schedulability of SBP significantly. 
However, one important downside is the \textit{performance interference} among multiple inference executions concurrently running on a GPU.
When the temporal sharing is exclusively used, one model inference is executed at a time, monopolizing the GPU resource during the execution.
However, inference executions running simultaneously may affect the performance of co-running inferences.
One common cause of such performance interference from co-running GPU tasks is the bandwidth of external memory, but other contentions on on-chip resource may engender performance degradation.
\begin{figure}[t]
	\begin{center}
		\includegraphics[width=1.0\linewidth]{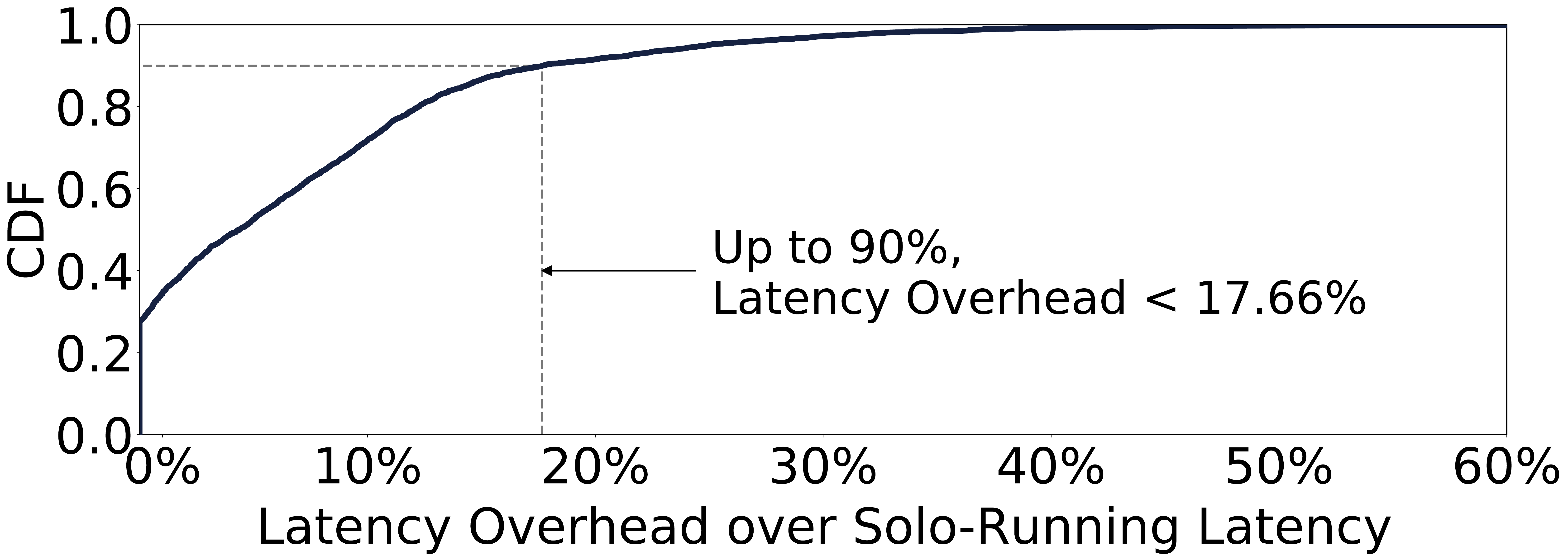}
	\end{center}
	\caption{Cumulative distribution of latency overhead when the pairs of inference executions are consolidated on a GPU.}
	\label{fig:interference_dist}
\end{figure}
%

To identify the interference effects, we perform an additional preliminary experimental study using a set of synthesized scenarios. 
We populate the pairs of two models drawn from the used five ML models (i.e., $_5C_2$ = 10) and associate the pairs with five different batch sizes (i.e., 2, 4, 8, 16, 32) to create 250 pairs in total. 
We also partition a GPU into two gpu-lets using various ratios: (2:8), (4:6), (5:5), (6:4), and (8:2). 
Then, we map the synthesized pairs to the different gpu-let pairs to observe the interference effects in various settings. 
Figure~\ref{fig:interference_dist} exhibits the cumulative distribution function (CDF) of latency overhead due to the consolidated inference executions, in comparison with the case where the models are run independently. 
As noted in the figure, for 90\% of the scenarios, the interference-induced overhead is lower than 18\%, which is modest. 
However, the CDF reports the long tail, which suggests the interference effect could be severe in certain circumstances. 
Thus, the interference may cause the scheduling decisions to be significantly incorrect, when the interfered executions produce latencies that are largely off from the expected range. 
Motivated by the insight, we devise a scheduling mechanism, which models the interference behaviors and leverage it to make the scheduling decisions robuster. 




\begin{table}[]
        \centering
        \small

        \setlength\tabcolsep{5.5pt}
        \begin{tabular}{ccccc}
                \hline
                \multirow{2}{*}{\textbf{Features}} &\textbf{Multi} & \textbf{Temporal} &\textbf{Spatial} &\textbf{Inter} \\
        &\textbf{Model} & \textbf{Scheduling} &\textbf{Scheduling} &\textbf{-ference} \\

                \hline
                Clipper~\cite{nsdi:clipper}     & \redx & \greencheck & \redx & \redx \\

                MArk~\cite{atc:mark}                    & \redx & \redx & \redx & \redx \\

                INFaaS~\cite{atc:infaas}                & \greencheck & \greencheck     & \redx & \redx\\

                Nexus~\cite{sosp:nexus}         & \greencheck & \greencheck     & \redx & \redx \\

                GSLICE~\cite{socc:gslice}               & \greencheck & \redx   & \greencheck & \redx \\

                \Name                                                   & \greencheck & \greencheck     & \greencheck & \greencheck \\
                \hline
        \end{tabular}
        \caption{Comparison with prior works related to GPU ML inference serving}
        \label{tab:prior_work}
\end{table}

\subsection{Comparison to the Prior Work}

Table~\ref{tab:prior_work} summarizes the advantages of gpu-let over the prior ML inference frameworks.
Gpu-let provides the scheduling of multiple heterogeneous models with both temporal and spatial resource 
scheduling. In addition, it considers the potential interference among partitions in the same GPU, with
interference modeling. Gpu-let is the only approach which consider all scheduling aspects.

\section{Design}
\label{sec:design}


\subsection{Overview}
This paper devises a scheduling scheme for multi-tenant ML inference serving, which aims to assign the incoming inference requests to the minimal number of GPUs, by leveraging both temporal and spatial sharing and maximizing the resource utilization.
Since the scheduling decisions need to be modified as the incoming request rates change, the scheduler is invoked periodically and the rescheduling is triggered if the current schedules produce SLO violations or the assigned resource is being underutilized.
%



\begin{figure}[t]
	\centering
	\includegraphics[width=\linewidth]{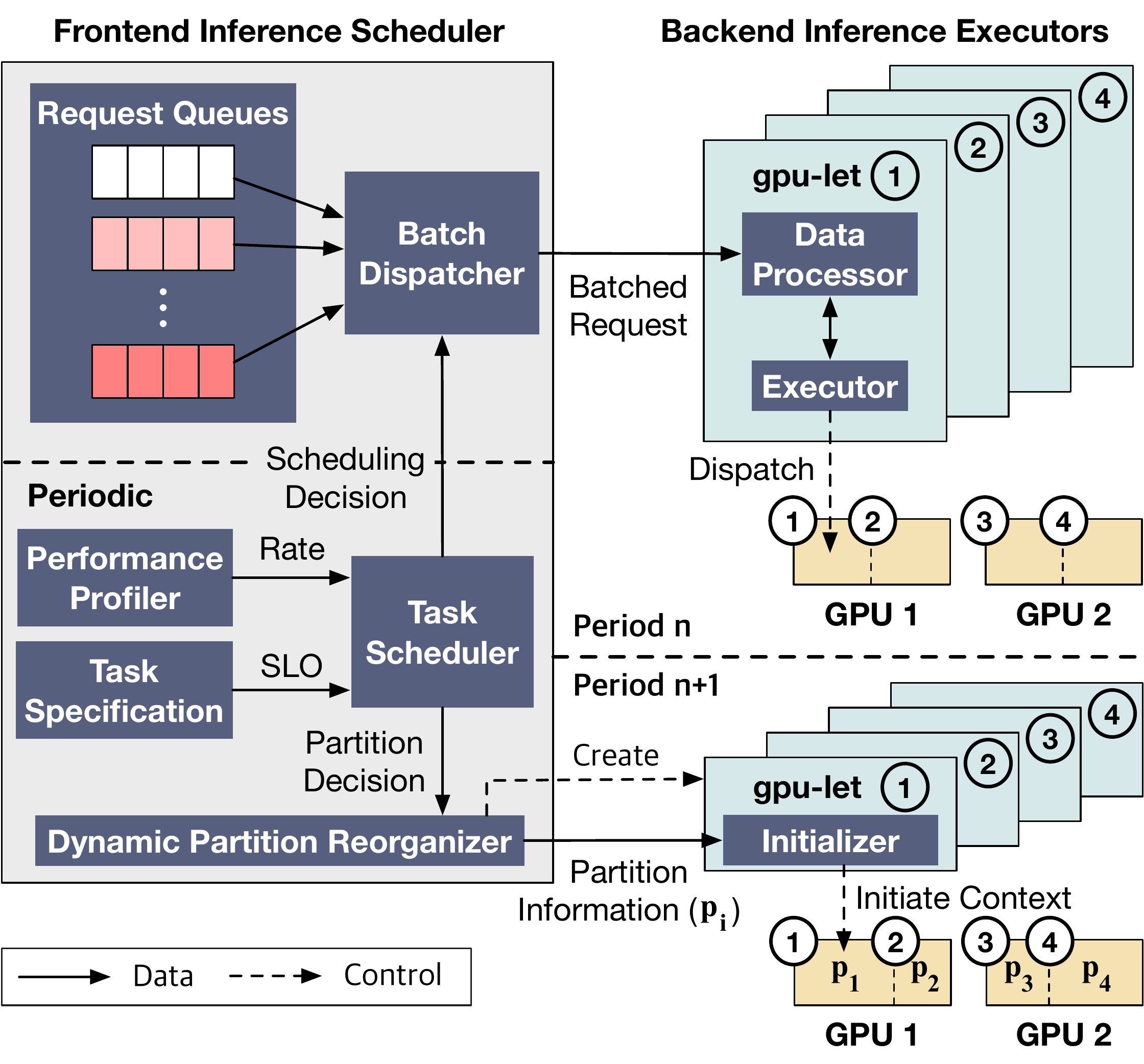}
	\caption{Overview of proposed scheduling scheme.}
        \vspace{0.1in}
	\label{fig:overview}
\end{figure}

Figure~\ref{fig:overview} presents the overall organization. 
%
%
The scheduler exploits the performance profiler, which not only monitors the incoming request rates, but also collects the performance statistics at offline, which are needed to make scheduling decisions (e.g., inference latency for a pair of batch size and gpulet size).
The user-specified task specification provides the model-specific SLO target. 
At runtime, the task scheduler is periodically triggered to check if the rescheduling is required. 
If yes, the scheduler makes two types of decisions: 1) the size of gpulets, and 2) the model assignments to the gpulets.
Based on the partitioning decisions, the dynamic partition reorganizer prepare the gpulets on the GPUs so that they can serve the inference request in the next period. 
The scheduling period is empirically determined based on the GPU partitioning latency so that the overhead for partitioning is hidden by the scheduling window.

%

\subsection{Challenge: Prohibitive Scheduling Search Space Size}
%
%
%
%

The core challenge in devising the scheduling algorithm is that the scheduling decision is dependent on various variables that are dependent to each other. 
How you partition the GPUs depends on the assigned model and its batch size, and determines the resource utilization, which is an optimization objective of the scheduler. 
Meanwhile, the batch size is dependent on the SLO constraints, which is the other objective. 
Therefore, the optimal scheduling decision would sit on the sweet spot in the search space built upon the three dimensions, 1) GPU partitioning, 2) batching, and 3) SLO guaranteeing, which create a huge search space. 
%

%
We formally analyze the space complexity of our scheduling problem below. 
Let $P$ be the number of GPU partitioning strategies on a GPU, $N$ be the number of GPUs to schedule, and $M$ be the number of models to serve. 
Each GPU can have $P$ possible \names so there are total $P^N$ possible strategies to partition $N$ GPUs.
Then, the $M$ models can be placed on the \names, possibly having all the $M$ models on a single \name. 
Since we need to check if the consolidation of multiple models violates the SLO, we must evaluate the total $M^2$ model placements per \name to assess the schedulability. 
As we have $NP$ \names on the system, the possible mappings of $M$ models to the \names is $NPM^2$. 
In summary, the space complexity of scheduling problem is as follows:

%

\vspace{0.1in}
\begin{center}	
	Search Space =  $\mathcal{O}$($P^NNPM^2$) 
\end{center}
\vspace{0.1in}

As the search space is prohibitively large, it is impractical to exhaustively search and pick an optimal solution, which extracts the best throughput from the minimal number of GPUs while guaranteeing SLO.  
Therefore, we deploy a heuristic, greedy approach, which effectively reduces the search space by estimating and allocating \names incrementally.

\subsection{Scheduling Algorithm}

In this section, we elaborate the scheduling algorithm. 
Table~\ref{table:notations} lists the variable notations used to describe the algorithm. 

\begin{table}
	\caption{Definition of variables for Elastic Partitioning.}
	\small
	\vspace{-2ex}
	\label{table:notations}
	\begin{center}
		\centering
		\begin{tabular}{cl}
			\hline
			\textbf{Name} & \textbf{Description} \\
			\hline
			$p$ & Partition size \\ 
			$b$ & Batch size \\ 
			$rate_i$ & Incoming rate of model \textit{i} \\
			$L$($b$,$p$) & Latency function of \textit{b} and \textit{p}\\
			$intf$ & Interference overhead function \\
			$SLO_i$ & SLO (in latency) of model \textit{i} \\
			$gpulet$.$size$ &  Actual partition size of $gpu\_let$  \\
			$alloc\_gpulets$ & Set of $gpu\_let$ with allocated model  \\
			$remain\_gpulets$ & Set of remaining $gpu\_let$ in system \\

			\hline
			
		\end{tabular}
	\end{center}
\end{table}





 
\setlength{\textfloatsep}{1.0pt}
\let\oldnl\nl
\SetKwComment{Comment}{$\triangleright$\ }{}
\newcommand{\nonl}{\renewcommand{\nl}{\let\nl\oldnl}}
\begin{algorithm}[t!]
	\small
	\LinesNumbered
	\SetAlgoLined
	\DontPrintSemicolon

	\nonl \textsc{ElasticPartitioning}($L$($b$,$p$), $intf$, $SLO$):
	


	\For(\tcp*[h]{If reorganization is required}){\emph{each period} }{

		Reset $remain\_gpulets$
				
		Sort all model by $rate_m$ in descending order
		
		$alloc\_gpulets$ $\leftarrow$ $\emptyset$

		\For{\emph{each model} $m$ }{
			
			$incoming\_rate$ $\leftarrow$ $rate_m$
			
			$assigned\_rate$ $\leftarrow$ 0
			
			\While{$incoming\_rate$ > $assigned\_rate$}{
				
				$p_{eff}$ $\leftarrow$ \textsc{MaxEfficientPartition}()
				
				$p_{req}$ $\leftarrow$ \textsc{MinRequiredPartition}( $rate_m$ )
			
				$p_{ideal}$ $\leftarrow$ \textsc{Min}($p_{eff}$, $p_{req}$)
			
			
				$gpulet$ $\leftarrow$ \textsc{FindBestFit}($p_{ideal}$, $SLO_m$, $intf$)
			
				$alloc\_gpulets \leftarrow alloc\_gpulets  \;\; \oplus $ \{$gpulet$\}
				
				$remain\_gpulets \leftarrow remain\_gpulets  \;\; \ominus $ \{$gpulet$\}
				
				$assigned\_rate$\ +=\ rate assigned to $gpulet$
			
			}

		}
		Apply $solution$ to system
	}		

\nonl \textsc{FindBestFit}($p\_{ideal}$, $SLO_m$, $intf$):


Sort all $gpulet$ by $gpulet.size$ in ascending order

\For{$gpulet$ \emph{in} $remain\_gpulets$}
{
	\If{$gpulet$.$size \geq p_{ideal}$}
	{
		\If{$gpulet$.$size$ == 100}
		{
			$gpulet_{ideal}$, $gpulet_{remain}$ $\leftarrow$ \textsc{Split}($gpulet$)
			
			$gpulet$ $\leftarrow$ $gpulet_{ideal}$
			
		}
		$b$ = $\mbox{argmax}_{k}$ $L(k, gpulet.size) \leq SLO$
		
		\If{$L(b, gpulet.size)$ + $intf \leq SLO$}
		{
			break
		}
	}
}
\For{$gpulet_{alloc}$ \emph{in} $alloc\_gpulets$}
{
	\If{$gpulet$ \emph{and} $gpulet_{alloc}$ \emph{are temporally sharable}}
	{
		$gpulet_{result}$ $\leftarrow$ \textsc{Merge}($gpulet$, $gpulet_{alloc}$)
		
		\textsc{RevertSplit}($gpulet$)
		
		return $gpulet_{result}$
	}
}
return $gpulet$

	\caption{\small Dynamic \name Scheduling Algorithm }
	\label{alg:elastic}
\end{algorithm}

\begin{figure}[t]
	\centering
	\includegraphics[width=0.75\linewidth]{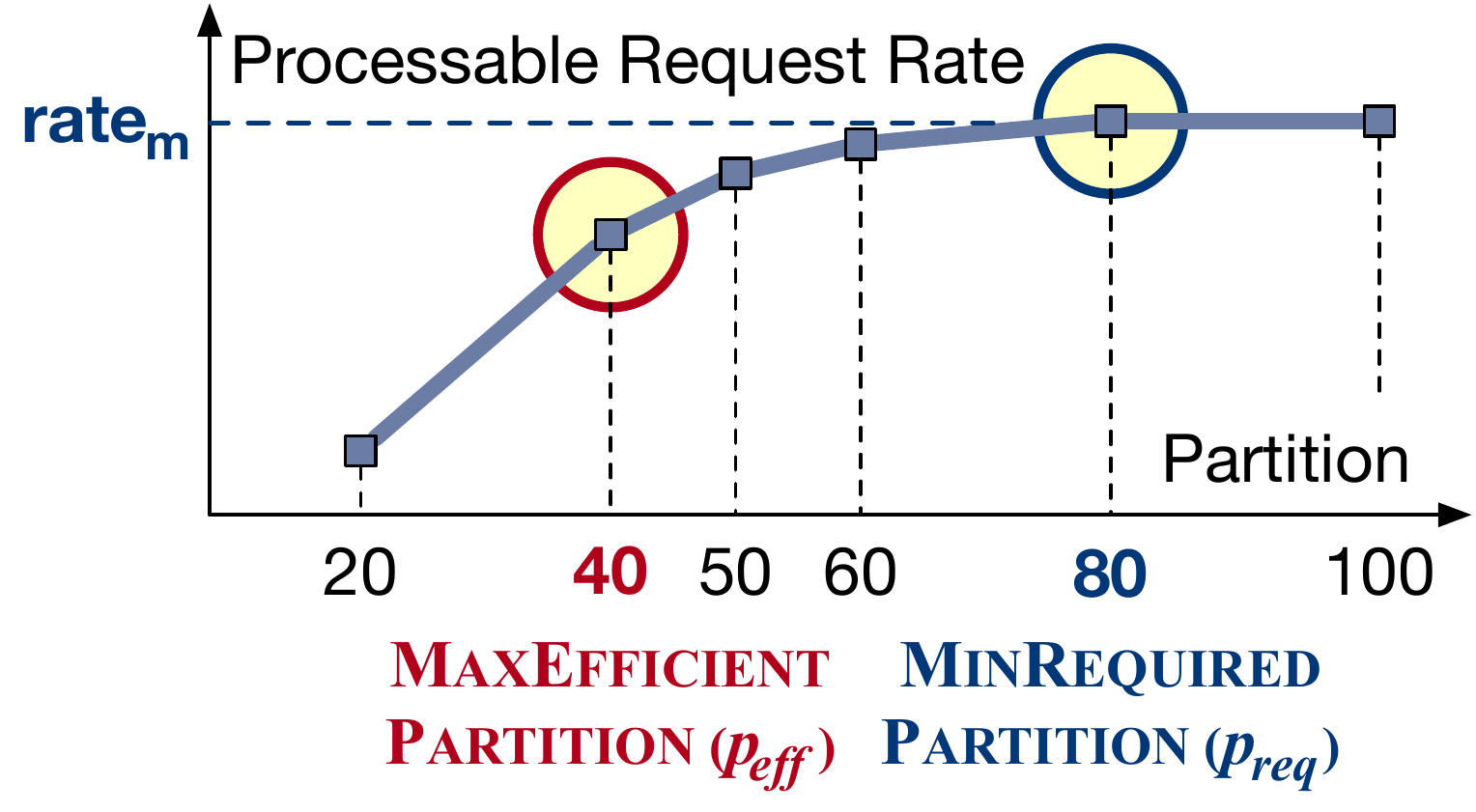}
	\caption{\textsc{MaxEfficientPart} and \textsc{MinRequiredPart}.}
	\label{fig:find_max_efficient_partition_with_MRP}
\end{figure}
 
\paragraph{Elastic partitioning:}
%
Algorithm~\ref{alg:elastic} describes the overall procedure of scheduling models to \names. 
\Alg receives the following input for each model: 1) $L$($b$,$p$): profiled execution latency of batch $b$ on partition size $p$, 2) $int\hspace{-0.15em}f$: interference function, and 3) $SLO$: per-model SLO. 
For every scheduling period, the server performs \alg for models that are running inference on the server if the request rates have been updated and thus a rescheduling is required ({\it line 1}). 
On one hand, if the request rates have been increased, the SLO violation could have happened so we need a reorganization. 
On the other hand, if the request rate have been decreased, the resource could have been underutilized and there might be opportunities to reduce the number of GPUs and save the resource. 
%
%
The incoming request rates of each model are tracked with an exponentially-weighted moving average and sorted in descending order ({\it line 2}). 
For each model $m$, the scheduler allocates one or more \names so that they can handle the entirety of request rate.

\paragraph{Determining ideal \name size:}
The algorithm first tries to maximize the system-wide throughput by choosing the ``ideal'' gpulet size between 1) the most cost-effective gpulet size ($p_{eff}$), and 2) the minimum required gpulet size to serve $rate_m$ while avoiding the SLO violation ($p_{req}$). 
To obtain the $p_{eff}$, the scheduler profiles at offline the affordable request rate for a given gpulet size. 
Figure~\ref{fig:find_max_efficient_partition_with_MRP} provides an example result of such profiling. 
In the curve, the knee, where the curvature has the local maximum, implies the most cost-effective sweet spot.
\textsc{MaxEfficientPartition} function calculates the curvature at the profiled gpulet size and uses the gpulet size at the knee as $p_{eff}$.
\textsc{MaxEfficientPartition} examines the minimum size of gpulet, $p_{req}$, which is necessary to eschew the SLO violation under the given request rate. 
%
The scheduler always picks the minimum of $p_{eff}$ and $p_{req}$. 
If $p_{req}$ is higher than $p_{eff}$, the unhandled request rate produced by the gap, $p_{req}$ - $p_{eff}$, is handled in the next iteration of the while loop ({\it line 8}). 

%

\paragraph{Incremental allocation with best-fit:}
After finding the ideal partition size $p_{ideal}$, \textsc{FindBestFit} performs a best-fit search and updates the server's remaining \names. 
%
First, the scheduler sorts \names by the partition size in an ascending order. 
The algorithm searches through $remain\_gpulets$ until a $gpulet,size$ is greater or equal to $p_{ideal}$. 
Since the \names are sorted in ascending order, the sweeping naturally guarantees the best-fit.
If the partition can be split, which means the \name chosen has a size of 100\%, the gpu is split by a function called \textsc{Split}. 
In the function, the \name is split into two \names, $gpulet_{ideal}$ and $gpulet_{remain}$, each with a size of $p_{ideal}$ and 100-$p_{ideal}$. 
Then, $gpulet_{remain}$ is added to $remain\_gpulets$.
The maximum batch size $b$ is decided and checked whether it can meet the SLO when there is the additional interference-induced overhead. 
After a $gpulet$ is chosen, the scheduler checks if the $gpulet$ is mergeable with any $gpulet_{alloc}$ through the temporal sharing since it can save the resource usage and in turn improve the resource utilization. 
If yes, the two are merged through \textsc{Merge}, which merges two \names. 
Finally, because $gpulet$ is now unused, the \textsc{RevertSplit} function reverts the partitioning happened for $gpulet$ and put it back to the $remain\_gpulets$ for future scheduling. 

\subsection{Modeling Interference}  \label{sec:interference_model}

A key challenge in the interference handling is to predict latency increases when multiple inferences are spatially partitioning a GPU with \names.
As shown in Figure~\ref{fig:interference_dist}, the interference effects are modest for the majority of consolidated executions, yet the overhead could be significant in infrequent cases.

In this study, we provide a simple interference-prediction model based on two key runtime behaviors of GPU executions.
The interference effects by spatial partitioning is commonly caused by the bandwidth consumption in internal data paths including the L2 cache,
and the external memory bandwidth. 
To find application behaviors correlated to the interference effects, we profiled the GPU with concurrent ML tasks with an NVIDIA tool (Nsight-compute).
Among various execution statistics, we found the {\it L2 utilization} and {\it DRAM bandwidth utilization} are the most relevant statistics
correlated to the interference.

Based on the observation, we build a linear model with the two parameters (L2 utilization and DRAM bandwidth utilization)  as follows: 

\begin{figure}[!t]
	\centering
	\includegraphics[width=0.9\linewidth]{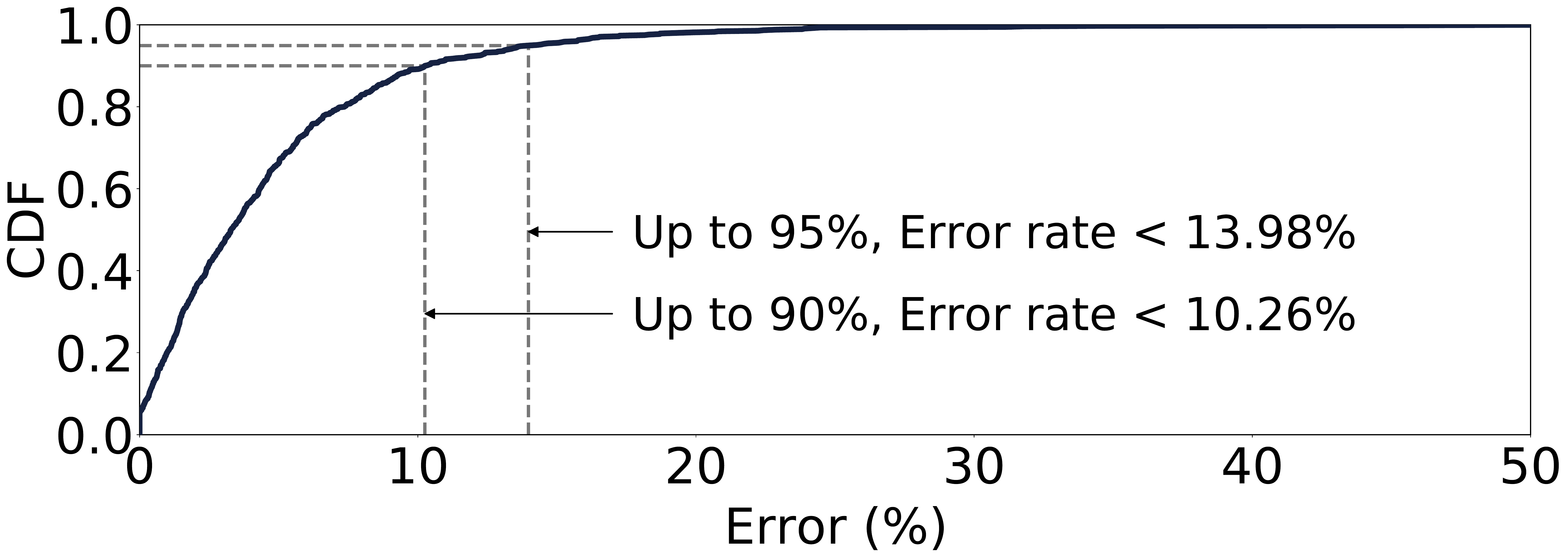}
	\caption{Cumulative distribution of relative error rate. Proposed analytical model can predict up-to 95\% of cases with less than 13.98 \% error rate.}
	\label{fig:interference_error}
\end{figure}

\vspace{0.1in}

{\small $interference\_factor = c1 \times l2_{m1}+c2 \times l2_{m2}+c3 \times mem_{m1}+c4 \times mem_{m2}+c5$ }
\vspace{0.1in}

$l2_{m1}$ and $l2_{m2}$ are L2 utilizations of model 1 and 2, when they are running alone with a given percentage of GPU resource. 
$mem_{m1}$ and $mem_{m2}$ are memory bandwidth consumption of model 1 and 2 with the same solo run.
Parameters (c1, c2, c3, c4, and c5) are searched with linear regression. 


We have profiled total 1,250 pairs (total 2,500 data) of inference interference and recorded how much interference each inference task has received.
Among 2,500 data, we have randomly selected 1,750 data of execution as training data and 750 data for validation. 
Figure~\ref{fig:interference_error} presents the cumulative distribution of the prediction error with our interference model.
The proposed model can predict up to 90\% of cases within 10.26\% error rate and up to 95\% if 13.98 \% of error is allowed.

\section{Implementation}
\paragraph{System software architecture and workflow:}
As illustrated in Figure~\ref{fig:overview}, our multi-GPU inference serving system constitutes a frontend inference scheduler and backend inference executors.
These software modules are instantiated as separate processes, which collaborate through interprocess communications.
The interprocess communication is developed using the Unix domain sockets. 
The software is developed in C++ and the approximate lines of code is 12.1K.
As the frontend scheduler receives the inference requests, it accumulates the requests for each model independently and forms a batch. 
The desired batch size per model is determined based on the scheduling decisions. 
When the desired size of request batch is formed or a duty-cycle is passed, the scheduler dispatches the batch to one of the backend inference executor processes, which is in charge of a gpu-let.
The backend inference executor processes are built based upon PyTorch, which initiates the inference execution for the given model and batched input. 
Once the results are returned from the GPUs, the backend process sends the results back to the scheduler process and in turn the scheduler serves the user application with the outputs. 
Below, we will delve into a few system modules and provide great details for better understanding.

\paragraph{Dynamic partition reorganizer:}
For our prototyped server, incoming rates are monitored and scheduled for a period of 20 seconds. 
The period was decided by an observation where reorganizing a GPUs partition takes approximately 10 to 15 seconds. 
The reorganization process includes spawning a new process with designated MPS resource, loading necessary kernels used by PyTorch, loading required models and warming up. 
The whole process is executed in background and has minimal effect on violating SLO as evaluated in Section~\ref{sec:evaluation}.


\paragraph{Backend inference executor:}
The inference executors receive the inference execution requests along with the corresponding model files and the batched input data from the scheduler.
The model file contains the model topology and parameters, which are archived as a pickled zip file using the .pt extension conventionally in PyTorch. 
The PyTorch runtime operating on the executors then initiate the inference executions using the given model and inputs on the gpu-let, leveraging the MPS-supported resource provisioning capabilities. 
To enable the users to control the resource provisioning feature, NVIDIA offers the MPS control daemon, which interfaces the hardware MPS features to the host software processes.
The MPS control daemon allows the users to set the active thread percentage, which controls the proportion of computing resource that will be enabled when the CUDA kernels run. 
%
%
%

%
%
%
%


\newcommand{\rulesep}{\unskip\ \vrule\ }
\section{Evaluation}
\label{sec:evaluation}
The proposed scheduling solution utilizes the otherwise wasted GPU resource, and enables multi-model GPU inference servers equipped with a fixed amount of resource to achieve higher \textit{throughput} and enhance \textit{schedulability}.
To examine the effectiveness, we use five different ML models that have different model topologies as well as different SLO latency requirements. 
We not only consider the inference executions of the five models as the independent requests, but also use mixes of these models to compose two different real-world application scenarios to evaluate more complicated scenarios. 
Firstly, we evaluate the maximum achievable throughput while not violating the SLO latency requirement, for the particularly chosen three model-level inference serving scenarios and two multi-model applications. 
Then, we show the efficacy of interference verification by looking into a few cases where considering overhead filters out request scenarios that would have produced non-negligible SLO violation rate. 
Next, we experiment how our solution can successfully adapt to rate fluctuation and use minimum \names as possible. 
Lastly, we compare schedulability of our solution to an ideal scheduling algorithm which exhaustively search all possible solutions.  
\subsection{Methodology}
\label{sec:method}
\paragraph{Inference serving system specifications:}
\begin{table}[t]
	\footnotesize
	\centering
			\begin{tabular}{ll}
			    \hline
			    \multicolumn{2}{c}{\textbf{System Overview}} \\
				\hline
				\textbf{CPU} & 20-core, Xeon E5-2630 v4 \\
				\textbf{GPU} & 4 $\times$ RTX 2080 Ti \\
				\textbf{Memory Capacity} & 192 GB DRAM \\
				\textbf{Operating System} & Ubuntu 16.04.6  \\
				\textbf{CUDA} &  10.2 \\
				\textbf{NVIDIA Driver} & 440.64 \\
				\textbf{ML framework} & PyTorch 1.2 \\
				\hline
			    \multicolumn{2}{c}{\textbf{GPU Specification}} \\
				\hline
				\textbf{CUDA cores} & 4,352  \\
				\textbf{Memory Capacity} & 11 GB GDDR6 \\
				\textbf{Memory Bandwidth} &  616 GB/sec \\
				\hline
			\end{tabular}
                 	\caption{The evaluated system specifications.}
			\label{tab:system-specifications}
\end{table}
Table~\ref{tab:system-specifications} provides the detailed descriptions of evaluated inference system and used GPU specifications. 
We use a multi-GPU inference server, which is equipped with a Xeon E5-2630 v4 CPU, four NVIDIA RTX 2080 Ti GPUs, and 192 GB of host DRAM memory. 
Note that the RTX 2080 Ti is a GPU card in the NVIDIA's Turing product line, which is a newer generation than the Volta and offers the post-Volta MPS capabilities. 
The table also provides the versions of used operating system, CUDA, and NVIDIA drivers. 
For ML inference executions, we use PyTorch, which is developed in a relatively programmer-friendly manner that facilitates the reasoning about the performance implications of scheduling decisions.  
%
%
We have a separate server to generate the inference requests, which is network-connected to the inference server, imitating the inference serving system architecture. 
The network bandwidth between the servers is 40 Gbps.
\paragraph{Baseline scheduling algorithms:}
For our baseline, we have ported the Squishy bin-packing(\abb{SBP}) algorithm ( from Nexus~\cite{sosp:nexus}) and a guided version of self-tuning algorithm \abb{guided self-tuning} introduced in GSLICE \cite{socc:gslice} on our prototype server. 
The original self-tuning algorithm dynamically searches batch size and suitable GPU partition size during runtime. Hence, it is unfair to compare it to a scheduling scheme which utilizes profiled information acquired beforehand. For fair comparison, we provide profiled batch execution latency and the optimal partition computed for models when evaluating \mbox{\abb{guided self-tuning}}.
Also, we have prepared two versions of our proposed algorithm in order to evaluate how effective our proposed interference model is. One does not consider interference when scheduling (\abb{gpulet}) and the other takes overhead caused by interference into consideration (\abb{gpulet+int}).

We do not provide a direct comparison to Nexus~\cite{sosp:nexus} due to the following reasons: 1) Nexus deploys optimizations which are orthogonal to our work, 2) several benchmarks that Nexus used in evaluation were not interoperable with our prototype server, such as models not supported by PyTorch.
However, we deploy the same video processing models that Nexus have used to evaluate their system and show how spatially partitioning GPUs can further enhance performance.
%

%
%

%
\paragraph{DNN models:}
\begin{table}[t]
	\small
	\centering
	\begin{tabular}{llc}
		\hline
		\multicolumn{1}{c}{\textbf{Model}} & \multicolumn{1}{c}{\textbf{Input Data (Dimension)}} & \multicolumn{1}{c}{\textbf{SLO (ms)}} \\
		\hline
		GoogLeNet (\abb{goo}) & Imagenet (3x224x224)) & 44  \\
		LeNet (\abb{le}) & MNIST (1x28x28)& 5 \\
		ResNet50 (\abb{res}) & Imagenet (3x224x224) & 95 \\
		SSD-MobileNet (\abb{ssd}) & Camera Data (3x300x300) & 136 \\
		VGG-16 (\abb{vgg})  &Imagenet (3x224x224)  & 130 \\
		\hline
	\end{tabular}
	\caption{List of ML models used in the evaluation.}
	\label{tab:ml-models}
\end{table}
Table~\ref{tab:ml-models} shows the list of five ML models used in our evaluation.
The models have a wide spectrum of model size and topologies, which lead to produce significant disparities among the inference latencies on the same GPU system. 
The last column, \texttt{SLO (ms)}, presents the per-model SLO latency constraint, which is set by doubling the solo execution latency of each model in our proposed system.  
The solo execution latency is determined depending on the model-inherent computational complexities and batch size.
Therefore, the larger batch size we use, the longer the SLO target latency becomes, but we use the batch size of 32, since using a larger value than 32 for the batch size engenders the SLO target latency unrealistically long.

\paragraph{Real-world multi-model applications:}
There are two real-world multi-model applications used in the evaluation, namely \abb{game} and \abb{traffic}.
These two applications are first introduced in Nexus~\cite{sosp:nexus} but we were unable to use the open-source implementations due to their incompatibility issues with PyTorch.
Thus, we develop the applications by ourselves, referring to application scenario descriptions provided by the paper. 
\begin{figure}[t]
	\centering
	\includegraphics[width=0.7\linewidth]{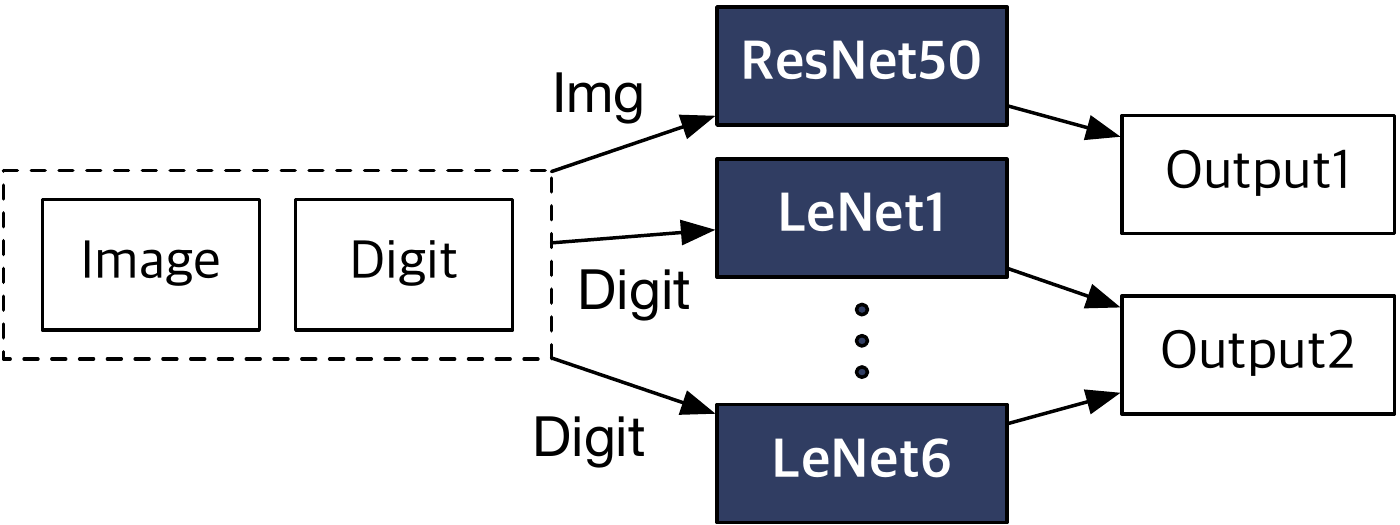}
	\caption{Dataflow of the multi-model scenario, `game'.}
	\label{fig:game}
\end{figure}
\begin{figure}[t]
	\centering
	\includegraphics[width=\linewidth]{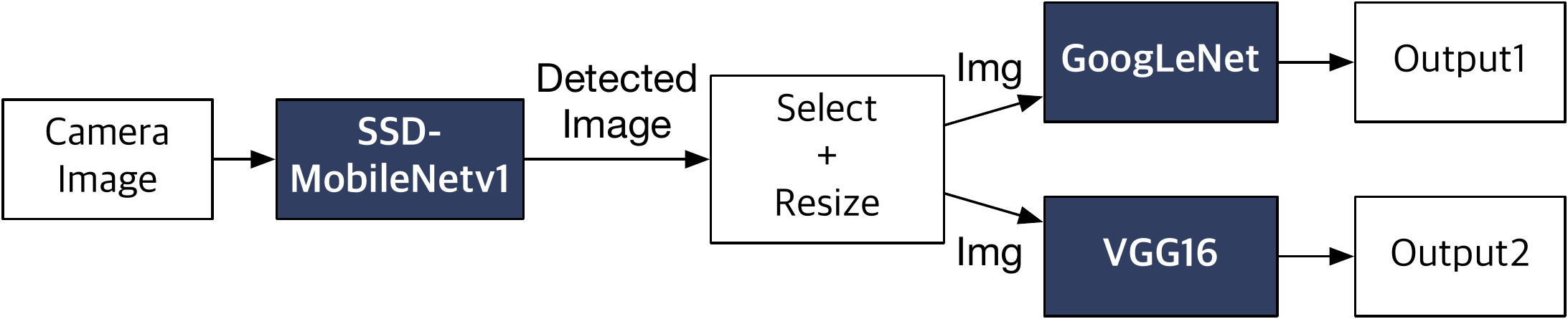}
	\caption{Dataflow of the multi-model scenario `traffic'.}
	\label{fig:traffic}
\end{figure}

Figure~\ref{fig:game} and Figure~\ref{fig:traffic} delineate the detailed dataflow graphs of the applications that contain the used ML models as well as the input/output data. 
The \textit{game} application analyzes the digits and image from the streamed video games in parallel, which consists of seven model inferences, the six of which are the LeNet inferences for digit recognition, while the rest is a ResNet-50 inference for image recognition.
The \textit{traffic} application is a traffic surveillance analysis, which has two phases. 
At the first phase, the application takes the camera image as the input and performs the object detection using the MobileNet-based SSD model.
Then, at the next stage, the application uses two different image recognition models, GoogLeNet and VGG-16, to recognize two different types of objects, respectively.  
The SLO latency is set as 95 ms and 136 ms for game and traffic, respectively. 
Each SLO latency is calculated by doubling the longest model inference latency.
\paragraph{Deeper look into the three particular request scenarios:}
\begin{table}[t]
	\small
	\centering
	\begin{tabular}{M{1cm}m{3.3cm}m{0.25cm}m{0.25cm}m{0.25cm}m{0.25cm}m{0.25cm}}
		\hline
		\multirow{2}{*}{\textbf{Name}} & \centering \multirow{2}{*}{\textbf{Description}} & \multicolumn{5}{c}{\textbf{Request Rate(req/s)}} \\
		&& le & goo & res  & ssd  & vgg  \\	\hline
		equal & All models with the identical request arrival rates  & 50 & 50  & 50   & 50   & 50   \\
		\hline 
		long-only &  Long-latency incurring models only with identical request arrival rates & 0  & 0   & 100 & 100 & 100 \\
		\hline
		short-skew & All models with the larger request arrival rates for small-latency incurring models & 100 & 100  & 100   & 50 & 50 \\
		\hline
	\end{tabular} 
	\caption{Three request scenarios, each of which represents a particular composition of multiple models where the request rate of each model is fixed as specified in the table. If a model has the zero requests per second, it means that the scenario does not involve the model's inference as a component.}
	\label{tab:particular-scenarios}
\end{table}
Among the aforementioned 1,023 request scenarios, we particularly choose three scenarios to take a deeper look. 
These three scenarios are characterized in a certain way, respectively. 
Table~\ref{tab:particular-scenarios} shows the details of the scenarios. 
The first scenario is \textit{equal}, which as the name implies, the five models have the same request arrival rates. 
The second scenario is \textit{long-only}, which is only consisted of relatively large models that have long latency. 
The last third scenario is \textit{short-skew}, which the most request models are small ones that incur short latency, while the requests for large ones exist in the mix. 
\paragraph{Request arrival rate:}
%
In order to evaluate our proposed system, we sample inter-arrival time for each model from a Poisson random distribution, based on a previous literature\cite{isca:treadmill} claiming that real-world request arrival intervals follow Poisson distribution. 

\paragraph{Runtime evaluation of request scenarios and applications:}
For a given scenario or application, we evaluate the scheduling decisions by actually deploying scheduling results on a prototype server and measuring the SLO violation rates. 
To embrace the unpredictable performance variations of the real system, we iterate the experiment three times for each scenario and application, and pick the median SLO violation rate. 
%
%
%

%
\subsection{Experimental Results}

\paragraph{Maximum achievable throughput comparison:}
We first evaluate the throughput implications of our schedulers.  
%
The \textit{maximum achievable throughput} is defined to be the request rates that the schedulers would return \texttt{Schedulable} while the produced schedules do not violate the SLO target when experimented on our prototype server. 
We obtain maximum achievable throughput of the schedulers by gradually increasing the request rate until SLO violation happens. 
\begin{figure}[t]
	
	\begin{center}
		\includegraphics[width=1.0\linewidth]{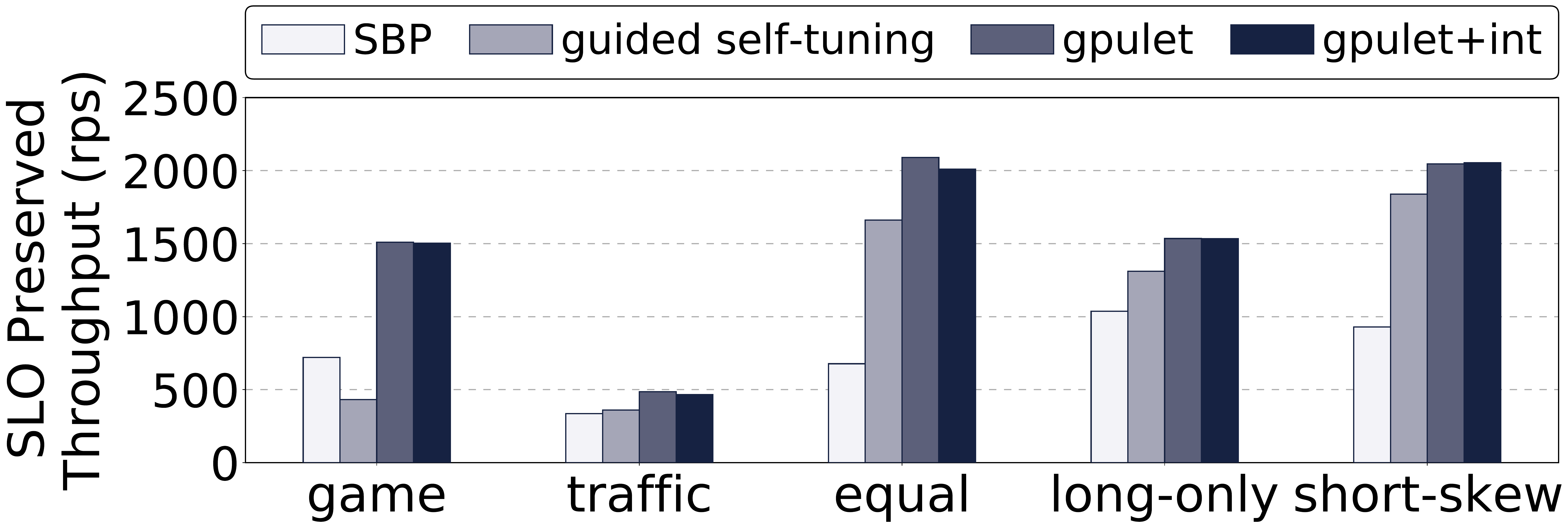}
	\end{center}
	\vspace{-1ex}
	\caption{ Maximum achievable throughput of the two multi-model applications (\abb{game} and \abb{traffic}) and three particularly chosen request scenarios (\abb{equal}, \abb{long-only}, and \abb{short-skew}).}
    \vspace{0.1in}
	\label{fig:slo-5models}
\end{figure}

Figure~\ref{fig:slo-5models} reports the maximum achievable throughput for the two multi-model applications and three particularly chosen request scenarios, when the four different scheduling algorithms are used. 
For all cases, our \abb{gpulet} and \abb{gpulet+int} schedulers offer an average 106.0\% and 102.6\% higher throughput than the \abb{SBP} baseline, respectively. 
\abb{gpulet} provides 3.4\% better throughput on average. The difference in throughput is caused by the conservative decisions made by the scheduler when additionally considering interference. However, we argue that such caution is necessary since a scheduler must be able to guarantee SLO at all times instead of maximizing throughput.
%
%
%

\abb{gpulet+int} also outperforms \abb{guided self-tuning} for every scenario with an average 74.8\% higher throughput.
The reason why \abb{guided self-tuning} shows such under-performance on \abb{game} is because \abb{game} consists of many short models(LeNet) and a single large model(ResNet50), making it more favorable for temporal sharing. ResNet50 received a 100\% \name due to suboptimal partitioning. Hence, the advantages of partitioning a GPU was minimized and disadvantages of not being able to support temporal sharing became significant.

Note that the reported throughput improvement is achievable merely through the MPS features already available in the most server-class GPUs and scheduling optimization in software, using exactly the same GPU machine.    
Thus, by making use of otherwise wasted GPU resource, the proposed scheduling scheme would be able to virtually offer cost savings for the ML inference service providers.  
For instance, \abb{gpulet+int} achieves 1502 req/s throughput for \abb{game} while \abb{SBP} does 720 req/s, utilizing the identical physical system, which can be translated into 47.9\% effective cost saving ($=\{1 - \frac{720}{1502}\} \times 100$).

%
%
%

\paragraph{Evaluation of meeting SLO:}
%
%
%
We take a deeper look into a certain level of maximum achievable request rate for the two multi-model application scenarios and three request scenarios to evaluate the efficacy of verifying interference. 
In order to evaluate and verify the system's capability to guarantee SLO we measure the percentage of requests that have violated the SLO, while counting dropped tasks also as SLO violating cases. 
We measured SLO violation by gradually increasing rates until both \abb{gpulet} and \abb{gpulet+int} consider the rate not schedulable, but only show the results of maximum rate for brevity. 

\begin{figure}[t]
	\begin{center}
		\includegraphics[width=1.0\linewidth]{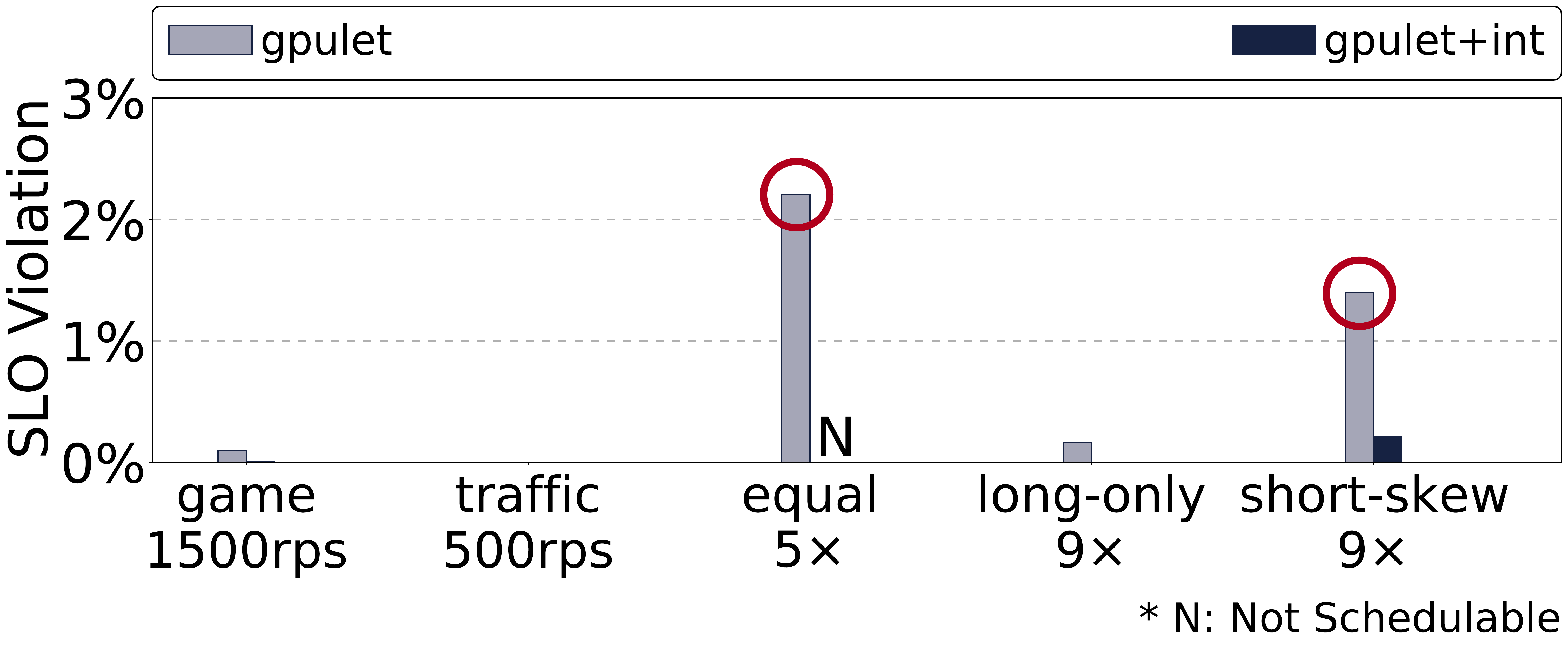}
	\end{center}
	\vspace{-2ex}
	\caption{SLO violation rates of two multi-model applications and three request scenarios. Rates were gradually increased until both \abb{gpulet} and \abb{gpulet+int} concluded the rate to be \textit{Not Schedulable}. The rate used for three request scenarios are denoted as \textit{x}$\times$ which is the factor of request rate increase ratio compared to the default rate listed in Table~\ref{tab:particular-scenarios}.}
	\vspace{1ex}
	\label{fig:slo-violation-3models-exp}
\end{figure}

Figure~\ref{fig:slo-violation-3models-exp} reports the SLO violation rate for each scenario and violation rates higher than 1\% are highlighted with the orange rounds.
\abb{gpulet}, which does not consider interference, shows violation rate higher than 1\% even for rates that it considered to be schedulable for \abb{equal} and \abb{short-skew}.
However, \abb{gpulet+int} successfully filters out such rate by either classifying such the rate as non-schedulable or successfully scheduling tasks to avoid SLO violation. 
%

%
%
%

%

%
%
 
%
%
%
%
%

\paragraph{Evaluation of adapting to rate fluctuation:}
To evaluate whether our prototype scheduler can successfully scale partitions for \names to accommodate fluctuating rates, we measure the performance of our scheduler while submitting inference queries with changing rates for all models presented in Table ~\ref{tab:ml-models}. 
The incoming rate for all models follow a Poisson distribution and each rate follows a unique trace different from one another in order to reproduce a realistic workload.

\begin{figure}[t]
  \centering
  \captionsetup[subfloat]{captionskip=0.3ex, labelformat=empty} 
  \subfloat[]{\includegraphics[width=1.0\linewidth]{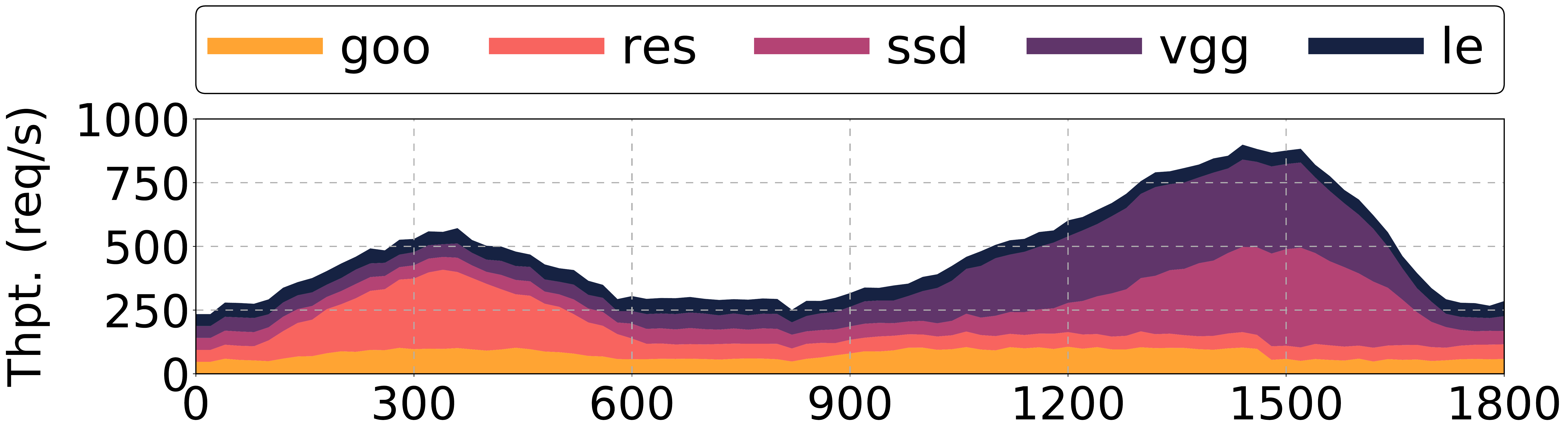}}\hfill
  \vspace{-4.5ex}
  \subfloat[]{\includegraphics[width=1.0\linewidth]{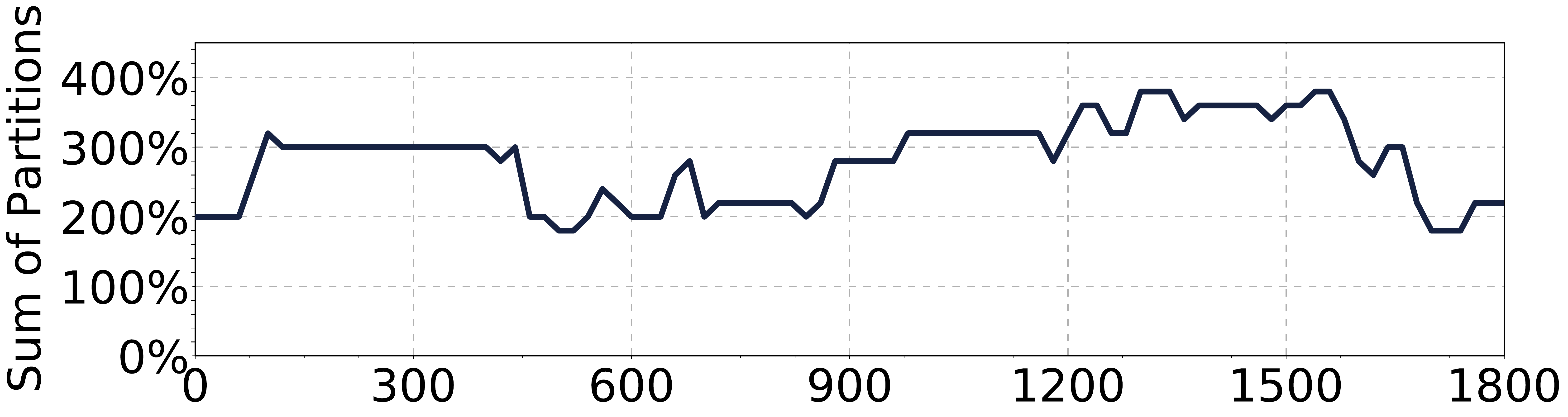}}\hfill
  \vspace{-4.5ex}
  \subfloat[]{\includegraphics[width=1.0\linewidth]{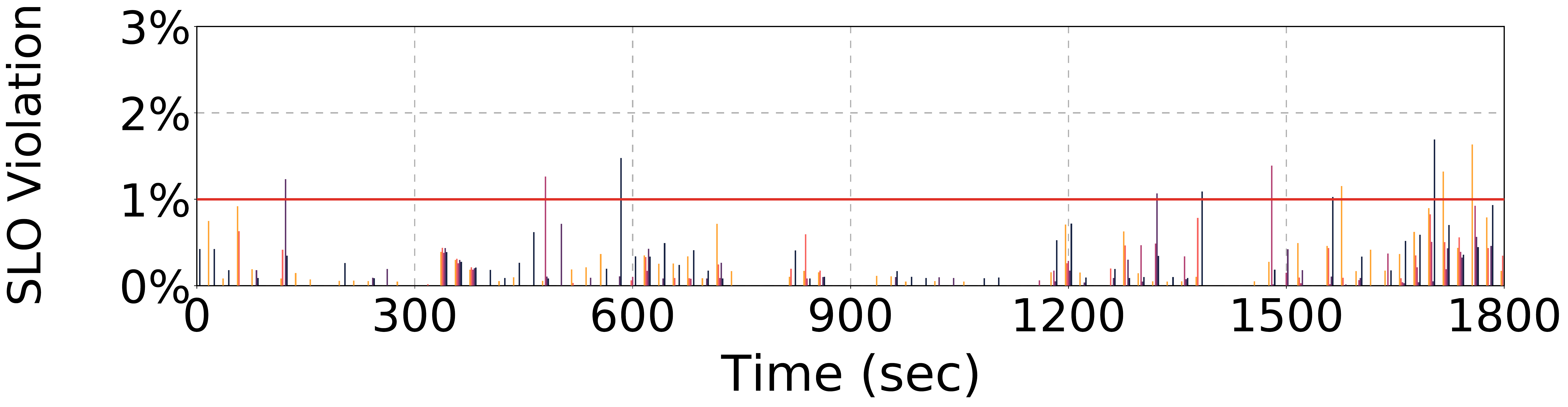}}\hfill
  \vspace{-2ex}
  \caption{Throughput, sum of utilized \name, SLO violation(\%) of each model for a 1,200 second window.}
  \label{fig:eval-5models}
\end{figure}

Figure ~\ref{fig:eval-5models} reports how our scheduling framework performed for a 1800 second window. The top graph shows a stacked graph of accumulated throughput of each model, the next graph sum of scheduled \name sizes and the last one SLO violation for 20 second periods.
Between 0 and 600 sec, the rate gradually increases and decreases to its initial rate. 
As the rate rises, our proposed scheduler successfully allocates more partitions to \names to preserve SLO. 
When the rate decreases, the sum of utilized partitions also decreases by reorganizing partitions. The following wave starting from 900 sec rises to a higher peak but our scheduler also successfully adjusts partitions. Although there are occasional SLO violations due to errors when predicting rates, note that the number of violated requests only accounts for 0.14\% of total requests.

Our proposed scheduler can successfully adjust to varying incoming rates by readjusting the size of each \name.

\paragraph{Comparison to ideal scheduler:} 
We evaluate the scheduling capability of \alg by comparing the scheduling results produced from an ideal scheduler. 
To produce the model-level inference request scenarios, we use the same methodology described in Section~\ref{sec:perf-impl}, which populates a set of 1,023 possible scenarios. 
The ideal scheduler makes scheduling decisions by exhaustively trying all possible \name combinations available for a given set of partitions. 
For example, the ideal scheduler will search through a total of $4^4$ cases, for 4 GPUs which can be partitioned into 4 cases. The search continues until all cases are searched or a case produces a viable scheduling result for a given request scenario.
For fair comparison, \abb{ideal} scheduler uses the same set of partitions as \abb{gpulet+int}.
\begin{figure}[t]
	\begin{center}
		\includegraphics[width=1.0\linewidth]{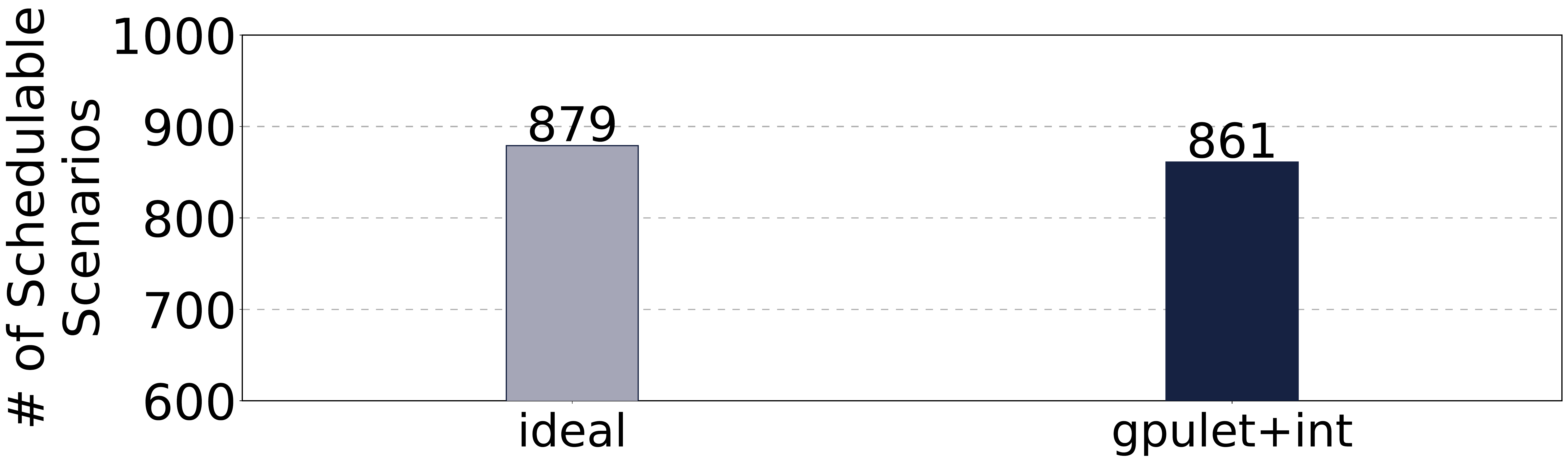}
	\end{center}
	\vspace{-2ex}
	\caption{Comparison between number of schedulable scenarios for ideal scheduler and gpulet+int scheduler.}
	\vspace{0.1in}
	\label{fig:ideal_1023}
\end{figure}
Figure ~\ref{fig:ideal_1023} compares the number of scenarios classified as \texttt{Schedulable} by each scheduler. 
\abb{gpulet+int} can schedule 18 less cases compared to \abb{ideal}, which is just 1.8\% of total 1023 cases.
\begin{figure}[t]
	\begin{center}
		\includegraphics[width=1.0\linewidth]{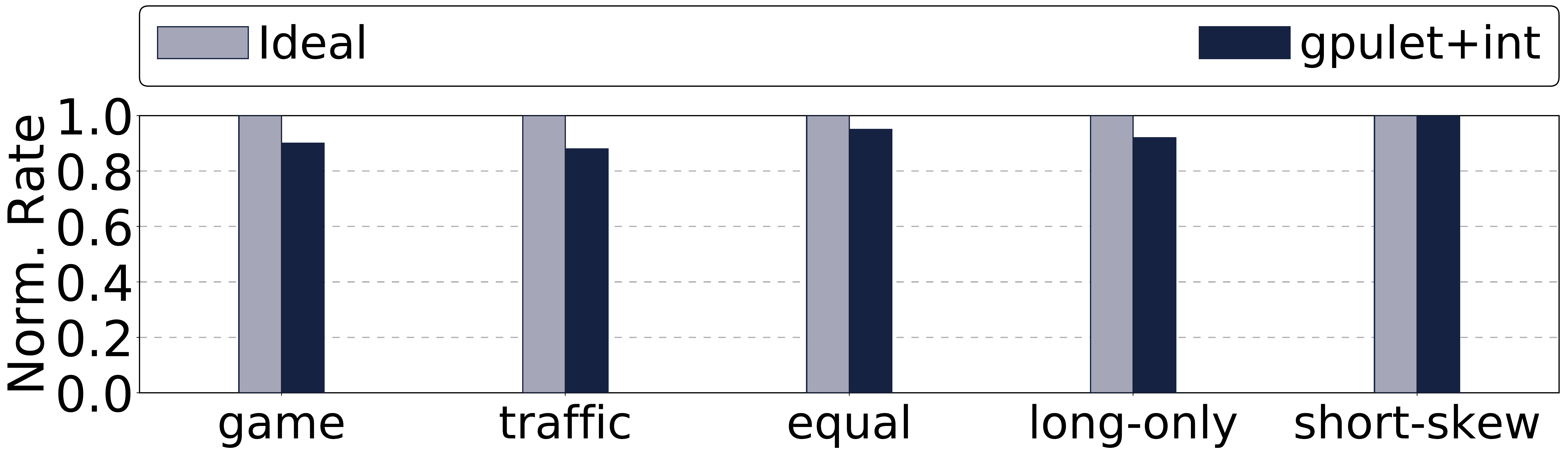}
	\end{center}
	\vspace{-2ex}
	\caption{Comparison of normalized maximum schedulable rates of real-world multi-model benchmarks and three chosen request scenarios.}
	\vspace{0.1in}
	\label{fig:ideal_macrobenchmark}
\end{figure}

Figure ~\ref{fig:ideal_macrobenchmark} reports the maximum schedulable rate of each multi-model scenario.  
All rates are normalized to the maximum rate which \abb{ideal} can provide. 
\abb{gpulet+int} can achieve an average 92.3\% maximum rate of the ideal throughput, with \abb{traffic} having the lowest rate of 87.7\%. 
%

\section{Related Work}

\paragraph{Machine learning service platforms.}
A wide variety of computer systems and researches have been proposed to improve quality of machine learning services\cite{atc:mark, aws:sagemaker, nsdi:clipper, vldb:rafiki, isca:djinn, osdi:tensorflow, nips:tensorflow_serving, nips:dynamic_space_time_scheduling, nvidia:tensorRT, OSDI:clockwork, sosp:nexus, socc:gslice, atc:infaas}. 
Nexus\cite{sosp:nexus} is a serving platform designed for DNN-based video analysis for GPU clusters. 
%
GSLICE \cite{socc:gslice} is a GPU-based inference serving platform which boosts performance by spatially sharing GPUs and hiding reorganization cost. 
 Although our work did not cover training, past researches inspired our work with schedulers for optimizing GPU resource\cite{eurosys:optimus, osdi:gandiva, isca:djinn, eurosys:geeps, atc:poseidon}.
%
%
Another related research direction focus on how to ease the burden of deployment and optimization for machine learning across various platforms \cite{osdi:TVM, osdi:ray, osdi:pretzel, nips:tensorflow_serving,  osdi:TVM, osdi:ray, osdi:pretzel, NIPS:nimble, nips:tensorflow_serving}. 

\paragraph{Interference estimation.}
Precise estimation of interference has been a key issue for high performance computing. Bubble-up\cite{micro:bubble_up} and bubble-flux\cite{isca:bubble_flux} models an application's sensitivity to cache and fits a sensitivity curve to predict performance.  Han {\it et al.}\cite{asplos:interference_management} extends using sensitivity cure to distributed computing where interference can propagate among processes. 
Baymax\cite{asplos:baymax} and Prophet\cite{asplos:prophet} models concurrent task execution behavior for non-preemptive accelerators. 

\paragraph{Multi-tenancy in Accelerator.}
NVIDIA GPUs have special HW/SW support for providing multi-tenancy to users. Such support include Multi Process Service (MPS)\cite{nvidia:MPS} with resource provisioning capabilities and  Multi Instance GPU (MIG) \cite{nvidia:MIG} with spatially partitioned HW. Similarly, AMD announced MxGPU \cite{amd:mxgpu} to process multiple instances.
Another field of study memory-hierarchy with multi-tenancy supports \cite{hpca:multi-tenancy-dws,hpca:multi-tenancy-dws}.  Choi {\it et al.} \cite{hpca:lazy-batching} proposes fine-grain batching scheme. PREMA \cite{hpca:prema} proposed time-multiplexing solution with preemption and Planaria \cite{micro:planaria} supports multi-tenancy by partitioning processing elements.

\section{Conclusion}

This study investigated an SLO-aware ML inference server design. It identified that
common ML model executions cannot fully utilize huge GPU compute resources when
their batch sizes are limited to meet the response time bound set by their SLOs.
Using spatial partitioning features, \names significantly improved throughput
of a multi-GPU configuration. The study shows ML inference servers
call for a new abstraction of GPU resources ({\it gpu-lets}), instead of using
conventional physical GPUs. 
The source code will become available. 

\section{Acknowledgements}
This work was supported the Institute
for Information \& communications Technology Promotion
(IITP2017-0-00466). The grant is funded by the Ministry
of Science and ICT, Korea.
\bibliographystyle{plain}
\bibliography{references}

\end{document}